\documentclass[apj,numberedappendix]{emulateapj}
\usepackage{apjfonts}
\journalinfo{The Astrophysical Journal, in press}
\submitted{Received 2004 June 14; accepted 2004 September 26}
\shorttitle{GCs with dark matter. I. Initial relaxation}
\shortauthors{Mashchenko and Sills}

\begin{document}

\title{Globular Clusters with Dark Matter Halos. I. Initial Relaxation}

\author{Sergey Mashchenko and Alison Sills}

\affil{Department of Physics and Astronomy, McMaster University,
Hamilton, ON, L8S 4M1, Canada; syam,asills@physics.mcmaster.ca}

\begin{abstract}
In a series of two papers, we test the primordial scenario of globular cluster
formation using results of high-resolutions $N$-body simulations. In this first
paper we study the initial relaxation of a stellar core inside a live dark
matter minihalo in the early universe. Our dark-matter dominated globular
clusters show features which are usually attributed to the action of the tidal
field of the host galaxy.  Among them are the presence of an apparent cutoff
(``tidal radius'') or of a ``break'' in the outer parts of the radial surface
brightness profile, and a flat line-of-sight velocity dispersion profile in the
outskirts of the cluster. The apparent mass-to-light ratios of our hybrid (stars
$+$ dark matter) globular clusters are very close to those of purely stellar
clusters. We suggest that additional observational evidence such as the presence
of obvious tidal tails is required to rule out the presence of significant
amounts of dark matter in present day globular clusters.
\end{abstract}


\section{INTRODUCTION}

The origin of globular clusters (GCs) is one of the important unsolved astrophysical
problems.  One can divide all proposed scenarios for GC formation into two large
categories: 1) in situ formation (when GCs are formed inside their present day
host galaxies), and 2) pregalactic formation (when GCs are formed in smaller
galaxies which later merge to become a part of the present day host galaxy). 

Pregalactic scenarios for GC formation can explain many
puzzling properties of GCs in the Milky Way and other galaxies. This is
especially true for so-called metal-poor GCs (with $[{\rm Fe/H}]\lesssim -1$),
which form a distinctive population in many galaxies. Many of their properties
are highly suggestive of their pregalactic origin: low metallicity, large age
comparable to the age of the universe, isotropic orbits with the
apocentric-to-pericentric distances ratio of $R_a/R_p\simeq 5.4\pm 3.7$
\citep[][based on data for 38 metal-poor Galactic GCs]{din99} which is consistent
with the orbits of dark matter (DM) subhalos ($R_a/R_p\sim 5\dots 6$) in
cosmological $\Lambda$CDM simulations, and a very weak or no
correlation with the properties of the host galaxy.

One very interesting variation of the pregalactic picture is the primordial
scenario of GC formation, which was first proposed by \citet{pee68}. In its
modern interpretation, the primordial scenario assumes that GCs formed at the
center of small DM minihalos in the early universe, before or during the
reionization of the universe, which is believed to have ended around the
redshift of $z=6$ \citep{bec01}. The primordial picture has been
considered by many authors, including \citet{pee84}, \citet{ros88}, \citet{pad97},
\citet{cen01}, \citet{bro02}, and \citet{bea03}.

It is usually assumed that during the reionization of the universe no star
formation can take place inside small DM halos with the virial temperature below
$\sim 10^4$~K (or virial mass less than $\sim 10^8$~$M_\odot$), as the gas in
such halos should escape the shallow gravitational potential after being heated
to $\sim 10^4$~K by the metagalactic Lyman continuum (LyC) background \citep{bar99}. Some
authors \citep*{cen01,ric02} suggested though, that the reionization of the
universe can actually trigger star formation in small DM minihalos through
different positive feedback mechanisms. In the scenario of \citet{cen01}, star
formation in small gas-rich minihalos is triggered by radially converging
radiation shock fronts caused by the external LyC radiation field. In the
cosmological simulations with radiative transfer of \citet{ric02}, small-halo
objects constitute the bulk of mass in stars until at least redshift $z\sim 10$
because of the increased non-equilibrium fraction of free electrons in front of
the cosmological H~{\sc ii} regions and inside the relic cosmological H~{\sc ii}
regions, which results in more efficient H$_2$ formation and hence cooling of
the gas.

Despite the fact that many authors considered different hydrodynamic and
radiative processes which can lead to formation of GC-like stellar clusters
inside DM minihalos, there has been no detailed study on what happens to such
hybrid (GC $+$ DM halo) object after the formation of stars from the point of
galactic dynamics. Fully consistent cosmological simulations of structure
formation in the universe (such as those of \citealt{ric02}), and even higher
resolution ``semi-consistent'' simulations (like in \citealt{bro02}) lack orders
of magnitude in spatial and mass resolution to be able to answer the following
questions: 1) How does the presence of DM modify the observable properties of GCs?
2) Are there observable features from which the presence of DM in a GC can be
inferred? 3) Will DM in ``hybrid GCs'' survive tidal
stripping during the hierarchical assemblage of substructure leading to the
formation of large galaxies such as Milky Way?

We try to answer the above questions in a series of two papers. In this first
paper, we address the first and the second questions, with the last question
being dealt with in the second paper \citep[][hereafter Paper~II]{mas04b}. Using
the $N$-body tree-code GADGET \citep*{SYW01}, we follow the relaxation of
initially non-equilibrium stellar clusters inside live DM minihalos around the
redshift of $z=7$. Our simulations are collisionless, and can be directly
compared with dynamically unevolved GCs; the impact of secular evolution (core
collapse) on our results will be explored in Paper~II.  DM halos have either
\citet[hereafter NFW]{NFW97} or
\citet{bur95} profiles, and have structural parameters taken from cosmological
simulations.  We use proto-GC model of \citet[hereafter MS04]{mas04a} to set up
the initial non-equilibrium configuration of stellar clusters inside DM
halos. In MS04 we showed that the collapse of homogeneous isothermal stellar
spheres leads to the formation of clusters with the surface brightness profiles
being very similar to those of dynamically young Galactic GCs. In this model, all
the observed correlations between structural and dynamic parameters of
Galactic GCs are accurately reproduced if the initial stellar density
$\rho_{i,*}$ and velocity dispersion $\sigma_{i,*}$ had the same universal
values: $\rho_{i,*}\simeq 14$~$M_\odot$~pc$^{-3}$ and $\sigma_{i,*}\simeq
1.91$~km~s$^{-1}$.

This paper is organized as follows. In Section~\ref{model} we describe our
method of simulating the evolution of a ``hybrid'' GC and list the physical and
numerical parameters of our models. In Section~\ref{results} we show the results
of the simulations. Finally, in Section~\ref{discussion} we discuss the results
and give our conclusions.

\section{MODEL}
\label{model}
\subsection{Initial Considerations}

In our model, GCs with a stellar mass $m_*$ are formed at the center of DM
halos with the virial mass $m_{\rm DM}$ at the redshift of $z=7$. We fix the
mass ratio $\chi\equiv m_*/m_{\rm DM}$ to 0.0088. The adopted value of $\chi$
is between the universal baryonic-to-DM density ratio $\Omega_b/\Omega_{\rm DM}=0.20$
\citep{spe03} and the fraction of baryons in GCs in the modern universe
of $\simeq 0.0025$ \citep{mcl99}. This leaves enough room for such effects as a
less than 100\% efficiency of star formation in proto-GCs, mass losses due to stellar
evolution (through supernovae and stellar winds), a decay of GC systems due to
dynamic evolution of the clusters in the presence of tidal fields, and any 
biased mechanism of GC formation (e.g., when GCs are formed only in DM halos
with a low specific angular momentum, like in \citealt{cen01}).

\subsection{DM Halos}

We consider a flat $\Lambda$CDM universe ($\Omega_\Lambda+\Omega_m=1$), with the following
values of the cosmological parameters: $\Omega_m=0.27$ and $H=71$~km~s$^{-1}$~Mpc$^{-1}$
\citep{spe03}. In a flat universe, the critical density can be written as

\begin{equation}
\rho_c(z)=\frac{3H^2}{8\pi G}\left[\Omega_m (1+z)^3+1-\Omega_m\right].
\end{equation}

\noindent The virial radius of a DM halo with a virial mass $m_{\rm DM}$ is then as follows:

\begin{equation}
\label{rvir}
r_{\rm vir}=\frac{1}{1+z}\left(\frac{2\,m_{\rm DM}G\,\Omega_m^z}{\Delta_c H^2\, \Omega_m}\right)^{1/3},
\end{equation}

\noindent where the spherical collapse overdensity, $\Delta_c$, and the matter
density of the universe in units of critical density at the redshift of $z$,
$\Omega_m^z$, are given by 

\begin{equation}
\Delta_c=18\pi^2+82x-39x^2
\end{equation}

\noindent and

\begin{equation}
\Omega_m^z=\left[1+\frac{1-\Omega_m}{\Omega_m\,(1+z)^3}\right]^{-1}
\end{equation}

\noindent \citep{bar01}. Here $x\equiv \Omega_m^z-1$.

We consider two types of DM density profiles: NFW halos and Burkert
halos. The NFW model describes reasonably well DM halos from cosmological CDM
simulations. It has a cuspy inner density profile with a slope of $\gamma=-1$,
and a steeper than isothermal outer density profile:

\begin{equation}
\rho_n(r) = \frac{\rho_{0,n}}{r/r_s\, (1+r/r_s)^2},
\end{equation}

\noindent where

\begin{equation}
\rho_{0,n}= \frac{m_{\rm DM}}{4\pi r_s^3}\left[\ln(1+c)-\frac{c}{1+c}\right]^{-1}.
\end{equation}

\noindent Here $r_s$ and $c\equiv r_{\rm vir}/r_s$ are the scale radius and concentration
of the halo. Burkert halos, on the other hand, have a flat core. Their outer
density profile slope is identical to NFW halos ($\gamma=-3$).  This model fits
well the rotational curves of disk galaxies \citep{bur95,sal00} and has the
following density profile:

\begin{equation}
\label{rho_b}
\rho_b(r) = \frac{\rho_{0,b}}{(1+r/r_s)\,[1+(r/r_s)^2]},
\end{equation}

\noindent where

\begin{equation}
\label{rho0_b}
\rho_{0,b}= \frac{m_{\rm DM}}{2\pi r_s^3}\left[\ln(1+c)+\frac12\ln(1+c^2)-\arctan c\right]^{-1}.
\end{equation}

The concentration $c$ of cosmological halos has a weak dependence on a virial mass
(with less massive halos being more concentrated on average). \citet*{ste02}
give the following expression for a concentration of low mass DM halos in
$\Lambda$CDM cosmological simulations at $z=0$, which was obtained from the
analysis of halos with the virial masses of $10^8\dots 10^{11}$~$M_\odot$: $c=27
(m_{\rm DM}/10^9 M_\odot)^{-0.08}$. Combining this expression with the result of \citet{bul01}
that the concentration scales with a redshift as $(1+z)^{-1}$, we obtained the following
formula for a concentration of low mass halos at different redshifts:

\begin{equation}
\label{eqc}
c=\frac{27}{1+z} \left(\frac{m_{\rm DM}}{10^9 M_\odot}\right)^{-0.08}.
\end{equation}

\noindent \citet{zhj03,zhm03} showed that CDM halos do not have concentrations
smaller than $\sim 3.5$, hence equation~(\ref{eqc}) becomes invalid for halos with
$c\lesssim 4$. \citet{ste02} demonstrated that equation~(\ref{eqc}) describes
very well concentrations of four dwarf disk galaxies from \citet{bur95} fitted 
by a Burkert profile, so it can be used for both NFW and Burkert halos.

\subsection{Stellar Cores}

The initial non-equilibrium configuration of proto-GCs was modeled after
MS04 as a homogeneous isothermal stellar sphere with isotropic
Maxwellian distribution of stellar velocities. Stellar clusters of different
mass $m_*$ initially had the same universal values of stellar density and
velocity dispersion: $\rho_{i,*}= 14$~$M_\odot$~pc$^{-3}$ and
$\sigma_{i,*}= 1.91$~km~s$^{-1}$ (MS04). As in MS04, we use
a mass parameter $\beta$ to describe different proto-GC models: 

\begin{equation}
m_*=10^\beta \sigma_{i,*}^3 \left(\frac{375}{4 \pi \rho_{i,*} G^3}\right)^{1/2},
\end{equation}

\noindent where $G$ is the gravitational constant. The connection between the
initial virial parameter $\nu$ and $\beta$ is $\nu \equiv 2K_*/W_* =
10^{-2\beta/3}$, where $K_*$ and $W_*$ are initial kinetic and potential
energies of the stellar cluster.


\begin{figure}
\plotone{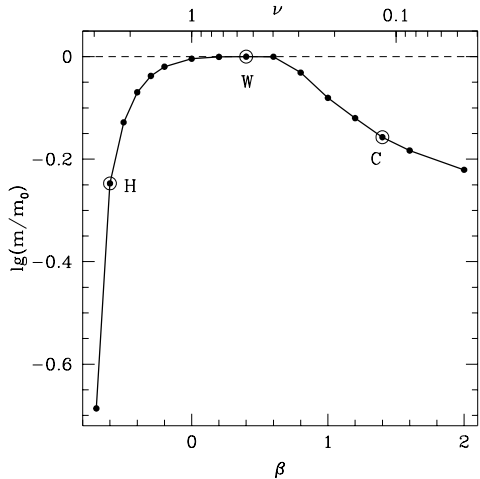}
\caption {Fraction of stars remaining gravitationally bound after the initial
relaxation phase in proto-GC models of MS04. Empty circles mark models with
$\beta=-0.6$ (``H''), $\beta=0.4$ (``W''), and $\beta=1.4$ (``C'').
\label{beta_m} }
\end{figure}


We use models from MS04 to plot in Figure~\ref{beta_m} a dependence of the
fraction of stars, which form a gravitationally bound cluster after the initial
relaxation phase of a proto-GC, on the mass parameter $\beta$ (or the virial
parameter $\nu$).  Please note that this is the case of no DM (bare stellar
cores). As one can see, there are three different regimes of the initial
relaxation phase: 1) ``Hot'' collapse ($-0.7\lesssim\beta\lesssim-0.35$ or
$3\gtrsim\nu\gtrsim 1.7$) results in a substantial loss of stars which initial
velocity was above the escape speed for the system; only the slowest moving
stars collapse to form a bound cluster. 2) ``Warm'' collapse
($-0.35\lesssim\beta\lesssim 0.85$ or $1.7\gtrsim\nu\gtrsim 0.27$) is very mild
and non-violent, and results in virtually no escapers. 3) ``Cold'' collapse
($\beta\gtrsim 0.85$ or $\nu\lesssim 0.27$) is violent, and leads to increasingly
larger fraction of escapers for colder systems. In the last case, the nature of
unbinding of stars is very different from the ``hot'' case. During a cold
collapse, the radial gravitational potential is wildly fluctuating. At the
intermediate stages of the relaxation, the central potential of the cluster
becomes much deeper than the final, relaxed value. As a result, a fraction of
stars are accelerated to speeds which will exceed the escape speed of the
relaxed cluster. Analysis shows that most of escapers are the stars that
initially were predominantly in the outskirts of the cluster and moving in the
outward direction.

These three regimes of the initial relaxation of proto-GCs are of very general
nature, and are not restricted to the GC formation scenario of MS04. To
illustrate that, let us write down an expression for the virial ratio of a
homogeneous isothermal sphere, which is applicable to both pre-starburst gas
phase of a forming GC, and the initial stellar configuration after the star
burst takes place:

\begin{equation}
\nu=\frac{5\langle V^2\rangle R}{3GM}.
\label{nueq}
\end{equation}

\noindent Here $M$ and $R$ are the mass and the radius of the system, and $\langle V^2\rangle$ 
is the mean-square speed of either gas molecules or stars. Assuming that
newly-born stars have the same velocity dispersion as star-forming gas,
equation~(\ref{nueq}) suggests three following physical scenarios resulting in
our ``hot'', ``warm'', and ``cold'' stellar configurations (in all scenarios we
start with an adiabatically contracting gas cloud of a Jeans mass, which
corresponds to $\nu_{\rm gas}=1$): 1) non-efficient stars formation with a
subsequent loss of the remaining gas leads to our ``hot'' case, 2) almost
100\%-efficient star formation results in the ``warm'' case, and 3)
runaway cooling of the whole cloud on the time scale shorter than the
free-fall time, leading to high-efficiency star formation, results in our ``cold''
case.

\begin{table*}
\caption{Physical parameters of the models\label{tab1}} 
\begin{center}
\begin{tabular}{cccccccccccccc}
\tableline
Model &$\beta$&$\nu_*$& $m_*$            &$r_*$ &$\tau_*$&$r_{*,\rm min}$&$m_{\rm DM}$& $c$  &$r_{\rm vir}$& $r_s$ &$r_{h,\rm DM}$&$\tau_{\rm DM}$& $S$ \\
      &       &       & $M_\odot$        & pc   & Myr    & pc            & $M_\odot$  &      & pc          & pc    & pc           & Myr           & \\
\tableline
W$_n$ &   0.4 & 0.54  & $8.8\times 10^4$ & 11.2 & 0.49   &    2.98       & $10^7$     & 4.88 &    885      &   181 &   395        &    52         & 4.7 \\
W$_b$ &   0.4 & 0.54  & $8.8\times 10^4$ & 11.2 & 0.49   &    2.98       & $10^7$     & 4.88 &    885      &   181 &   436        &    61         & 117 \\
C$_n$ &   1.4 & 0.12  & $8.8\times 10^5$ & 24.2 & 0.078  &    1.57       & $10^8$     & 4.06 &   1906      &   470 &   892        &    56         & 5.8 \\
C$_b$ &   1.4 & 0.12  & $8.8\times 10^5$ & 24.2 & 0.078  &    1.57       & $10^8$     & 4.06 &   1906      &   470 &   997        &    66         & 173\\
H$_n$ & --0.6 & 2.5   & $8.8\times 10^3$ & 5.21 & 17     &    4.14       & $10^6$     & 5.87 &    411      &    70 &   174        &    48         & 3.7 \\
H$_b$ & --0.6 & 2.5   & $8.8\times 10^3$ & 5.21 & 17     &    4.14       & $10^6$     & 5.87 &    411      &    70 &   190        &    55         & 78 \\
\tableline
\end{tabular}
\end{center}
\tablecomments{
%
Here $\nu_*$ and $r_*$ are the initial virial ratio (assuming,
that there is no DM) and radius of the stellar core; $\tau_*$ is the crossing
time at the half-mass radius for the relaxed, purely stellar models from MS04
rescaled to $\rho_{i,*}= 14$~$M_\odot$~pc$^{-3}$ and $\sigma_{i,*}=
1.91$~km~s$^{-1}$; $r_{*,\rm min}$ is the minimum attained half-mass radius for
relaxing stellar clusters from MS04 (rescaled to the above values of
$\rho_{i,*}$ and $\sigma_{i,*}$); $r_{h,\rm DM}$ and $\tau_{\rm DM}$ are the
half-mass radius and the crossing time at the half-mass radius for DM halos;
$S\equiv m_*/m_{\rm DM}(r_*)$ is the ratio of the stellar mass to the mass of DM
within the initial stellar radius $r_*$.
}
\end{table*}

\subsection{Physical Parameters of the Models}
\label{physical}

In this paper, we consider three following initial stellar configurations, which
can be considered to be typical ``hot'', ``warm'', and ``cold'' cases (see
Figure~\ref{beta_m}): $\beta=-0.6$ (model ``H'', stands for ``Hot''),
$\beta=0.4$ (model ``W'', stands for ``Warm''), and $\beta=1.4$ (model ``C'',
stands for ``Cold'').  We study relaxation of stellar cores in either NFW
(suffix ``$n$''; e.g., ``W$_n$'') or Burkert (suffix ``$b$'') DM halos, which
makes up a total of 6 different combinations of stellar cores and DM halos.  The
physical parameters of these 6 models are summarized in Table~\ref{tab1}.

The stellar masses $m_*$ in our models cover the whole range of GC masses (see
Table~\ref{tab1}).  The initial stellar radius $r_*$ is much smaller than the
scale radius of the DM halo $r_s$ in all the models. The masses of DM halos
$m_{\rm DM}$ range from $10^6$ to $10^8$~$M_\odot$. From Figure~6 of
\citet{bar01}, these DM halos correspond to $1-1.5\sigma$ fluctuations
collapsing at the redshift of $z=7$. As one can see, halos in this mass range
are populous enough at $z\sim 7$ to account for all observed GCs.


\begin{figure}
\plotone{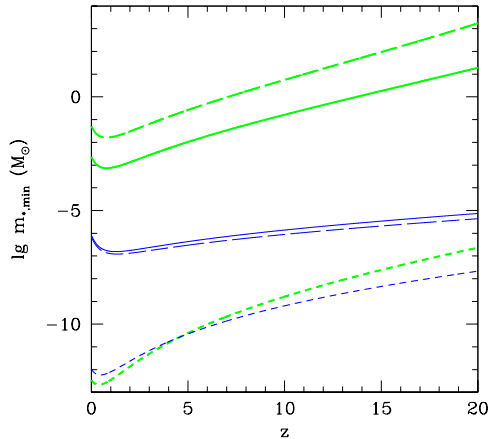}
\caption {
Redshift dependence of the minimum mass of a stellar core, with the universal
initial density $\rho_{i,*}= 14$~$M_\odot$~pc$^{-3}$, which dominates DM at the
center of a DM halo. Thick lines (colored green in electronic edition)
correspond to stellar cores in NFW halos, whereas thin lines (colored blue in
electronic edition) correspond to clusters in Burkert halos.  Long-dashed,
solid, and short-dashed lines correspond to $\chi=0.0025, 0.0088, 0.2$.
\label{M_min} }
\end{figure}


In all our models, stars initially dominate DM in the core ($S\equiv m_*/m_{\rm
DM}[r_*]>1$, see Table~\ref{tab1}), which can be considered as a desirable
property for a star formation to take place. It is instructive to check if the
above condition ($S>1$) holds for other plausible values of the GC formation
redshift $z$ and stellar mass fraction $\chi$. In Figure~\ref{M_min} we show the
redshift dependence of the minimum mass of a stellar core in a DM halo,
satisfying the condition $S\ge 1$, for a few fixed values of $\chi$ --- 0.0025,
0.0088 (our fiducial value), and 0.20. Both NFW (thick lines) and Burkert (thin
lines) cases are shown.  As one can see, only in the extreme case of an NFW halo
at $z=20$ with $\chi=0.0025$ fraction of total mass in stars, the minimum mass of a
dominant stellar core $m_{*,\rm min}=1.7\times 10^3$~$M_\odot$ is approaching
the mass range of GCs. The conclusion we arrive at is that, for all plausible
values of $z$ and $\chi$, stellar cores with the universal initial density
$\rho_{i,*}= 14$~$M_\odot$~pc$^{-3}$ and GC masses dominate DM at the center of
DM halos.

\subsection{Setting Up $N$-Body Models}

DM halos in our models are assumed to have an isotropic velocity dispersion
tensor.  To generate $N$-body realizations of our DM halos, we sample the
following probability density function (PDF) \citep{wid00}:

\begin{equation}
P(E,R)\propto R^2 (\Psi-E)^{1/2} F(E).
\label{pdfeq}
\end{equation}

\noindent Here $\Psi$ and $E$ are the relative dimensionless potential and particle
energy in units of $4\pi G \rho_{0,n} r_s^2$ (for NFW halos) and $\pi^2 G
\rho_{0,b} r_s^2$ (for Burkert halos), $F(E)$ is the phase-space distribution
function, and $R\equiv r/r_s$ is the dimensionless radial coordinate of the
particle.  The dimensionless potential $\Psi$ is equal 1 at the center and 0 in
infinity. For NFW halos, $\Psi=R^{-1}\ln(1+R)$. For Burkert halos, the
corresponding expression is given in Appendix~\ref{app1}
(eq.~[\ref{psibur}]). We use the analytic fitting formulae of
\citet{wid00} to calculate $F(E)$ for NFW model, and derive our own fitting
formulae for the Burkert profile (see Appendix~\ref{app1}). Our models are
truncated at the virial radius $r_{\rm vir}$ (eq.~[\ref{rvir}]): $F(E)=0$ for
$r>r_{\rm vir}$. As we will show in \S~\ref{warm_sect}, this truncation results
in an evolution in the outer DM density profile (with the profile becoming
steeper for $r\gg r_s$), which should not affect our results.

We sample the PDF (eq.~[\ref{pdfeq}]) in two steps: 1) A uniform random number
$x\in [0\dots 1]$ is generated. The radial distance $R$ is then obtained by
solving numerically one of the two following non-linear equations:

\begin{equation}
\ln (1+R)-R/(1+R) = x\,[\ln(1+c)-c/(1+c)]
\end{equation}

\noindent for the NFW profile and

\begin{eqnarray}
&\ln(1+R)+[\ln(1+R^2)]/2-\arctan R = &\nonumber\\
&= x\left[\ln(1+c)+[\ln(1+c^2)]/2-\arctan c\right]&
\end{eqnarray}

\noindent for the Burkert profile. 2). Given $R$, we can calculate $\Psi$.
Instead of equation~(\ref{pdfeq}), we can now sample the following PDF:

\begin{equation}
P(E)\propto \Pi\equiv \left(\frac{\Psi-E}{\Psi}\right)^{1/2} \frac{F(E)}{F(\Psi)}.
\end{equation}

\noindent As $F(E)$ is a monotonically increasing function and $0\le E\le \Psi$, 
the following inequality holds: $0\le \Pi\le 1$. Two uniform random numbers, $y\in
[0\dots 1]$ and $E\in [0\dots \Psi]$, are generated. If $y\le\Pi$, we accept
the value of the dimensionless energy $E$; otherwise, we go back to the previous
step (generation of $y$ and $E$). 

The velocity module $\upsilon$ can be obtained from the equation for the total
energy of a particle: $E=\upsilon^2/2+\Psi$.  Finally, two spherical coordinates
angles $\theta$ and $\varphi$ are generated for both radius and velocity vectors
using the following expressions: $\cos \theta =2t-1$ and $\varphi=2\pi u$. (Here
$t$ and $u$ are random numbers distributed uniformly between 0 and 1.)

The above method to generate $N$-body realizations of either NFW or Burkert halos
does not use a ``local Maxwellian approximation'' to assign velocities to
particles. Instead, it explicitly uses phase-space distribution functions, which
was shown to be a superior way of setting up $N$-body models
\citep{kaz04}.

Stellar cores are set up as homogeneous spheres at the center of DM halos:
$R=r_* x^{1/3}$, where $x\in[0\dots 1]$ is a uniformly distributed random
number.  We use equal mass stellar particles.  Velocities of the particles have
a Maxwellian distribution. The components of the radius and velocity vectors are
generated in the same fashion as for DM particles.

\subsection{Numerical Parameters of the Models}

\begin{table}
\caption{Numerical parameters of the models\label{tab2}} 
\begin{center}
\begin{tabular}{ccccccccc}
\tableline
Model & $N_*$ & $\epsilon_*$ &  $N_{\rm DM}$ & $\epsilon_{\rm DM}$ & $t_1$ & $t_2$         & $\Delta t_{\rm max}$ & Note\\
      &       &    pc        &               &        pc           & Myr   &     Myr       &       Myr            & \\
\tableline
W$_0$ & $10^5$&   0.15       &       \nodata &  \nodata            &  190  &     370       &    0.3               &1\\
W$_n$S& $10^4$&   0.30       &       \nodata &  \nodata            &   95  &     190       &    0.2               &2\\
W$_n$ & $10^4$&   0.30       &        $10^6$ &  1.5                &  200  &     250       &    0.2               &\nodata\\
W$_b$S& $10^4$&   0.30       &       \nodata &  \nodata            &   95  &     190       &    0.2               &2\\
W$_b$ & $10^4$&   0.30       &        $10^6$ &  1.5                &  150  &     250       &    0.2               &\nodata\\
\tableline

C$_0$ & $10^5$&   0.32       &       \nodata &  \nodata            &   56  &      92       &    0.3               &1\\
C$_n$ & $10^5$&   0.32       &$5\times 10^5$ &  4.7                &  130  &     190       &    0.2               &\nodata\\
C$_b$ & $10^5$&   0.32       &$5\times 10^5$ &  5.2                &   95  &     190       &    0.2               &\nodata\\
\tableline

H$_0$ & $10^5$&   0.069      &       \nodata &  \nodata            & 1300  &    1800       &    0.3               &1\\
H$_n$ & $10^4$&   0.15       &$5\times 10^5$ &  1.7                & 1500  &    3700       &    20                &\nodata\\
H$_b$ & $10^4$&   0.15       &$5\times 10^5$ &  1.8                & 1500  &    3700       &    20                &\nodata\\
\tableline
\end{tabular}
\end{center}
\tablecomments{
%
1) Models from MS04 rescaled to $\rho_{i,*}=
14$~$M_\odot$~pc$^{-3}$ and $\sigma_{i,*}= 1.91$~km~s$^{-1}$ (stars only, no
DM). 2) Static DM potential. Here $N_*$ and $N_{\rm DM}$ are the number of
stellar and DM particles; $\epsilon_*$ and $\epsilon_{\rm DM}$ are the softening
lengths for stars and DM; $t_1$, $t_2$, and $\Delta t_{\rm max}$ are the moment
of time when most of model parameters converge to their quasi-steady-state
values, the total evolution time, and the maximum value for individual
timesteps.
}
\end{table} 

In addition to 6 models from Table~\ref{tab1} which consist of a live DM halo
plus a stellar core, we also ran 2 more models with a static DM potential (marked
with the letter ``S'' at the end). The numerical parameters for these 8 models
are listed in Table~\ref{tab2}. In this table we also list parameters for the
three properly rescaled stars-only models from MS04 (H$_0$, W$_0$, and
C$_0$), which will be used as reference cases.

To run our models, we use a parallel version of the multistepping tree
code GADGET-1.1 \citep{SYW01}. The code allows us to handle separately
stellar and DM particles, with each species having a different softening length
--- $\epsilon_*$ and $\epsilon_{\rm DM}$, respectively.  We slightly modified
the code by introducing an optional static DM potential (either NFW, or
Burkert).

The values of the softening lengths were chosen to be comparable with the
initial average interparticle distance. We used the following expression for the
stellar particles: $\epsilon=0.77 r_h N^{-1/3}$ \citep{hay03}, where the
half-mass radius $r_h$ was measured at the initial moment of time (so
$r_h=r_*/2^{1/3}$). The same expression was used to calculate $\epsilon_{\rm
DM}$ for H$_{n,b}$ models, whereas for the models W$_{n,b}$ and C$_{n,b}$ we
used half of the value given by the equation of \citet{hay03} to achieve a
better spatial resolution within the stellar core.

\citet*{alp88} showed that collapsing homogeneous stellar spheres with 
a non-negligible velocity dispersion $\sigma_{i,*}$ do not experience a
fragmentation instability (and as a consequence preserve orbital angular
momentum of stars), if the following condition is met: $N_*\gtrsim [m_*/(r_*
\sigma_{i,*}^2)]^3$, where $N_*$ is the number of stars. For our models, this
adiabaticity criterion can be rewritten as

\begin{equation}
\label{eqbeta}
\log N_* \gtrsim 2\beta + 3\log 5.
\end{equation}

\noindent As we discussed in MS04, if real GCs indeed started off as isothermal
homogeneous stellar spheres, they should have collapsed adiabatically.  We also
showed that if both the condition in equation~(\ref{eqbeta}) and
$\epsilon_*\lesssim 0.25 r_{*,\rm min}$ are met, the results of simulations of
collapsing stellar spheres do not depend on the number of particles $N_*$ and
the softening length $\epsilon_*$. (Here $r_{*,\rm min}$ is the minimum
half-mass radius of the cluster during the collapse.) By comparing
Tables~\ref{tab1} and \ref{tab2} one can see that the values of $N_*$ and
$\epsilon_*$ we use in our simulations meet both above conditions.

The evolution time $t_2$ in our runs is at least 3 times longer than the
crossing time for DM $\tau_{\rm DM}$, and hundreds and even thousands times longer
than the crossing time for stellar particles $\tau_*$ (see Tables~\ref{tab1} and
\ref{tab2}). As we will see in the next section, this time is enough for stellar
and DM density profiles to converge. On the other hand, it is short enough
to avoid significant dynamic evolution in the stellar clusters caused by
encounters between individual particles.

The individual time steps in the simulations are equal to $(2\eta
\epsilon/a)^{1/2}$, where $a$ is the acceleration of a particle, and parameter
$\eta$ controls the accuracy of integration. We used a very conservative value
of $\eta=0.0025$ and set the maximum possible individual time step $\Delta t_{\rm max}$
value to either 20 or 0.2~Myr (see Table~\ref{tab2}). As a result, the accuracy
of integration was very high: $\Delta E_{\rm tot}\lesssim 0.06$\%, where $\Delta
E_{\rm tot}$ is the maximum deviation of the total energy of the system from its
initial value. One has to keep in mind though that numerical artifacts are the
most pronounced in the central densest area, where the frequency of strong
gravitational interaction between particles is the highest. We estimate the severity
of these effects by looking at the total energy conservation in our purely stellar
models from MS04: in models H$_0$, W$_0$, and C$_0$ the values of $\Delta E_{\rm tot}$
are $\sim 0.05$\%, $\sim 0.85\%$, and $\sim 0.8$\% (for the model W$_0$ we used a
larger value of $\eta=0.02$, hence the relatively large errors). As you can see,
even though the numerical artifacts in the dense stellar cluster  are more visible
than in the DM $+$ a stellar core case, the magnitude of these effects is still
reasonably low.

For completeness sake, we also give the values of other code parameters which
control the accuracy of simulations: ErrTolTheta=0.6,
TypeOfOpeningCriterion=1, ErrTolForceAcc=0.01, MaxNodeMove=0.05,
TreeUpdateFrequency=0.1, and DomainUpdateFrequency=0.2. (Please see the code
manual\footnote{\url{http://www.mpa-garching.mpg.de/gadget/}} for explanation
of these parameters). The code was compiled with the option DBMAX enabled,
which allowed it to use a very conservative node-opening criterion.

\section{RESULTS OF SIMULATIONS}
\label{results}
\subsection{General Remarks}

\begin{table*}
\caption{Parameters for relaxed stellar clusters\label{tab3}} 
\begin{center}
\footnotesize{
\begin{tabular}{ccccccccccccc}
\tableline
Model & $r_{h,*}$ & $\sigma_c$  &  $\rho_c$      &  $R_{h,*}$ &  $R_{hb}$ &  $\sigma_0$ & $\Sigma_V$ & $r_0$ & $f_0$      & $\Upsilon$  &  $r_\rho$ &  $r_m$\\
      &  pc       &  km~s$^{-1}$&$M_\odot$~pc$^3$&    pc      &    pc     & km~s$^{-1}$ &mag~arcsec$^{-2}$& pc    &       &$M_\odot$~$L_\odot^{-1}$&   pc      &  pc\\
\tableline
W$_0$ & 4.64 &      4.07     &   310              &  3.64     &  2.77     &   3.84   &18.71  &  3.01 & 0.226 & 1.43  &  \nodata  &  \nodata\\
W$_n$S& 4.51 &      4.36     &   360              &  3.50     &  2.63     &   4.11   &18.63  &  2.96 & 0.253 & 1.60  &  \nodata  &  \nodata\\
W$_n$ & 4.08 &      4.83     &   500              &  3.16     &  2.31     &   4.55   &18.39  &  2.78 & 0.279 & 1.80  &  7.74     &  19.7    \\
W$_b$S& 4.67 &      4.05     &   310              &  3.64     &  2.81     &   3.83   &18.73  &  2.98 & 0.238 & 1.43  &  \nodata  &  \nodata\\
W$_b$ & 4.65 &      4.12     &   350              &  3.61     &  2.54     &   3.89   &18.65  &  2.86 & 0.229 & 1.52  &  17.8     &  54.7    \\
\tableline

C$_0$ & 2.55 &      14.4     &   $1.3\times 10^4$ &  2.03     &  1.52     &   13.6   &15.31  &  1.64 & 0.241 & 1.43 &  \nodata  &  \nodata\\
C$_n$ & 4.50 &      14.3     &   $1.1\times 10^4$ &  3.54     &  1.55     &   13.4   &15.45  &  1.75 & 0.166 & 1.55 &  10.0     &  44.0    \\
C$_b$ & 4.50 &      14.4     &   $1.3\times 10^4$ &  3.52     &  1.49     &   13.5   &15.30  &  1.62 & 0.168 & 1.42 &  27.6     &  125    \\
\tableline

H$_0$ & 20.2 &      0.54     &   0.71             &  14.9     &  7.13     &   0.51   &24.28  &  8.29 & 0.154 & 1.64  &  \nodata  &  \nodata\\
H$_n$ & 5.47 &      1.57     &   25               &  4.20     &  2.79     &   1.56   &21.44  &  4.06 & 0.329 & 2.88  &  \nodata  &  \nodata\\
H$_b$ & 11.5 &      0.97     &   4.8              &  8.84     &  4.43     &   0.97   &22.70  &  5.71 & 0.212 & 2.23  &  10.4     &  18.2    \\
\tableline
\end{tabular}}
\end{center}
\tablecomments{
%
Here $r_{h,*}$, $\sigma_c$, and $\rho_c$ are half-mass radius, central velocity
dispersion, and central density of stellar clusters; $R_{h,*}$, $R_{hb}$,
$\sigma_0$, and $\Sigma_V$ are projected half-mass radius, half-brightness
radius (where surface brightness is equal to one half of the central
surface brightness), projected central dispersion, and central surface
brightness (see eq.~[\ref{eqSigma}]); $r_0$ and $f_0$ are the King core radius
$r_0=[9\sigma_c^2/(4\pi G \rho_c)]^{1/2}$ and the fraction of the total stellar
mass inside $r_0$; $\Upsilon$ is the apparent central mass-to-light ratio (see
eq.~[\ref{eqUpsilon}]); $r_\rho$ is the radius where density of DM and stars
becomes equal; $r_m$ is the radius where enclosed mass of DM becomes equal to
that of stars.  
}
\end{table*} 

\begin{figure*}
\plottwo{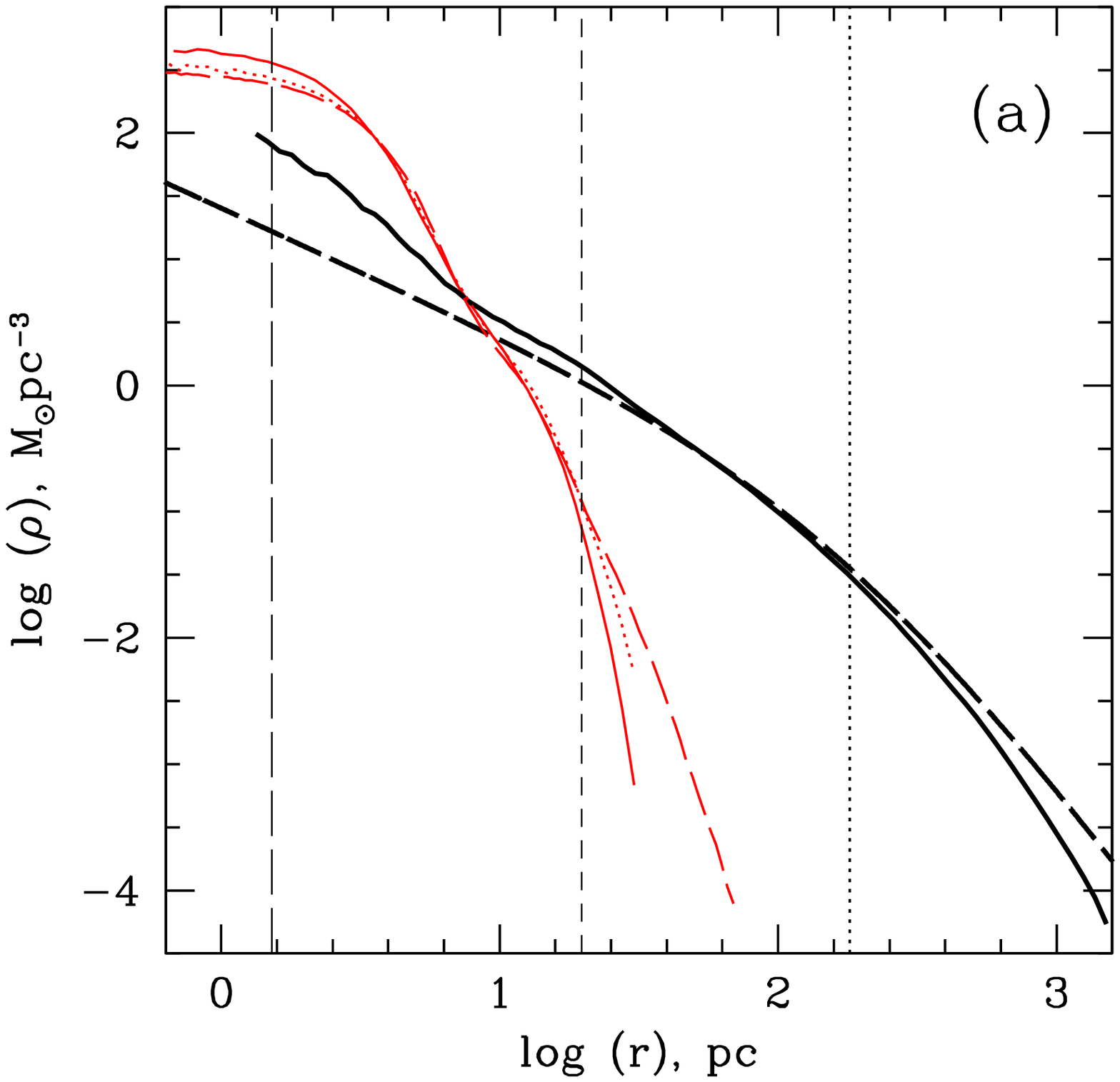}{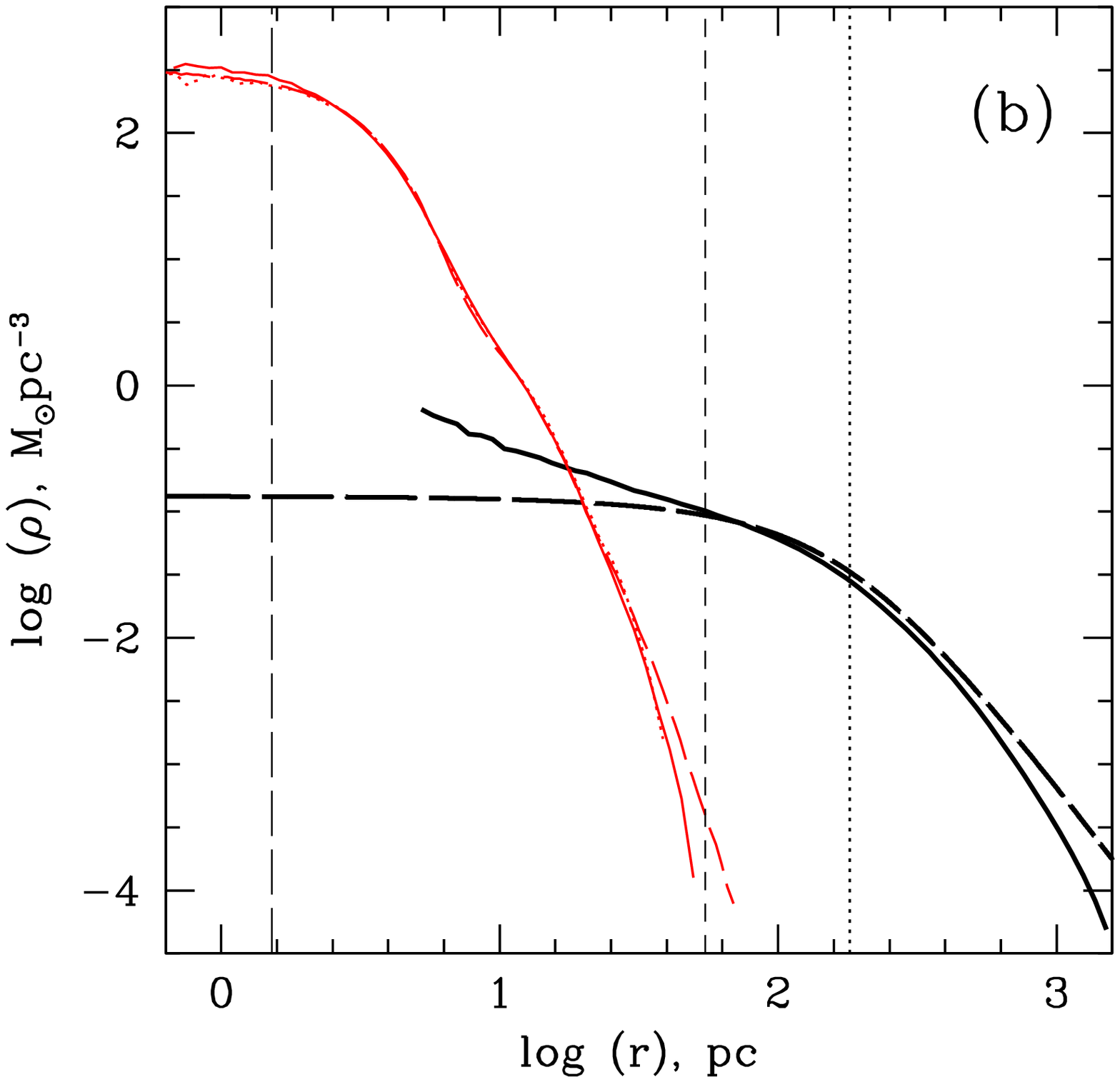}
\caption {
Radial density profiles for warm collapse models. (a) The case of NFW halo. (b)
The case of Burkert halo. Thick lines correspond to DM density; thin lines
(colored red in electronic edition) show stellar density. Solid lines correspond
to the cases of a live DM halo $+$ a stellar core (models W$_n$ and W$_b$),
long-dashed lines depict analytic DM profiles and a relaxed stellar cluster
profile in the absence of DM (model W$_0$), and dotted lines show stellar
density profiles for static DM models (W$_n$S and W$_b$S). Vertical long-dashed,
short-dashed, and dotted lines mark the values of $\epsilon_{\rm DM}$, $r_m$,
and $r_s$.
\label{rho_W} }
\end{figure*}

For each of our models, we generated 100--200 individual timeframe snapshots.
The last 20--60\% of the snapshots (corresponding to the time interval $t_1\dots
t_2$, see Table~\ref{tab2}) were used to calculate the properties of relaxed
models, including radial profiles and different stellar cluster parameters
listed in Table~\ref{tab3}. In all snapshots, we removed in an iterative manner
the escapers (particles whose velocity exceeds the escape velocity $V_{\rm
esc}=[-2\Phi]^{1/2}$, where $\Phi$ is the local value of the gravitational
potential). This procedure affected models from MS04 (H$_0$, W$_0$, and C$_0$),
and DM particles only in the models presented here.  One of the reasons for
removing escapers was to make the results of the simulations directly comparable
with the results from Paper~II where this procedure is applied to discount
particles stripped by tidal forces.

The radial profiles shown in this section were obtained by averaging the
corresponding profiles from individual snapshots in the time interval
$t_1\dots t_2$. For each profile, we only show the part which is sufficiently
converged (the dispersion between different snapshots is very small). For
projected quantities (such as surface brightness $\Sigma_V$ and 
line-of-sight velocity dispersion $\sigma_r$) we use the projection method
described in Appendix~\ref{app2} which produces much less noisy radial profiles
than a straightforward projection onto one plane or three orthogonal planes. This
is crucial for the analysis of the properties of the outer, low density parts
of stellar clusters, which is the main purpose of this study.

The virial ratio $\nu$ for our models is consistent with unity within the
measurement errors at the end of simulations. All the global model parameters
(listed in Table~\ref{tab3}) converge to their final values by
$t=t_2$. Similarly, radial density and velocity dispersion profiles converge to
their final form over a wide range of radial distances. We explicitly checked
that the evolution time $t_2$ is more than three crossing times at the largest
radius for stellar density profiles shown in Figures~\ref{rho_W}, \ref{rho_C},
and \ref{rho_H}.  All the above let us conclude that the results presented here
are collisionless steady-state solutions.

\subsection{Warm Collapse}
\label{warm_sect}

In the absence of DM, an isothermal homogeneous stellar sphere with the mass
parameter $\beta=0.4$ relaxes to its equilibrium state with virtually no stars
lost (see Figure~\ref{beta_m}). This makes it an interesting case to test the
result of \citet{pee84}, that a stellar cluster inside a static constant density
DM halo can acquire a radial density cutoff similar to that expected from the
action of tidal forces of the host galaxy, for the case of live DM halos with
cosmologically relevant density profiles and initially non-equilibrium stellar
cores.


\begin{figure*}
\plottwo{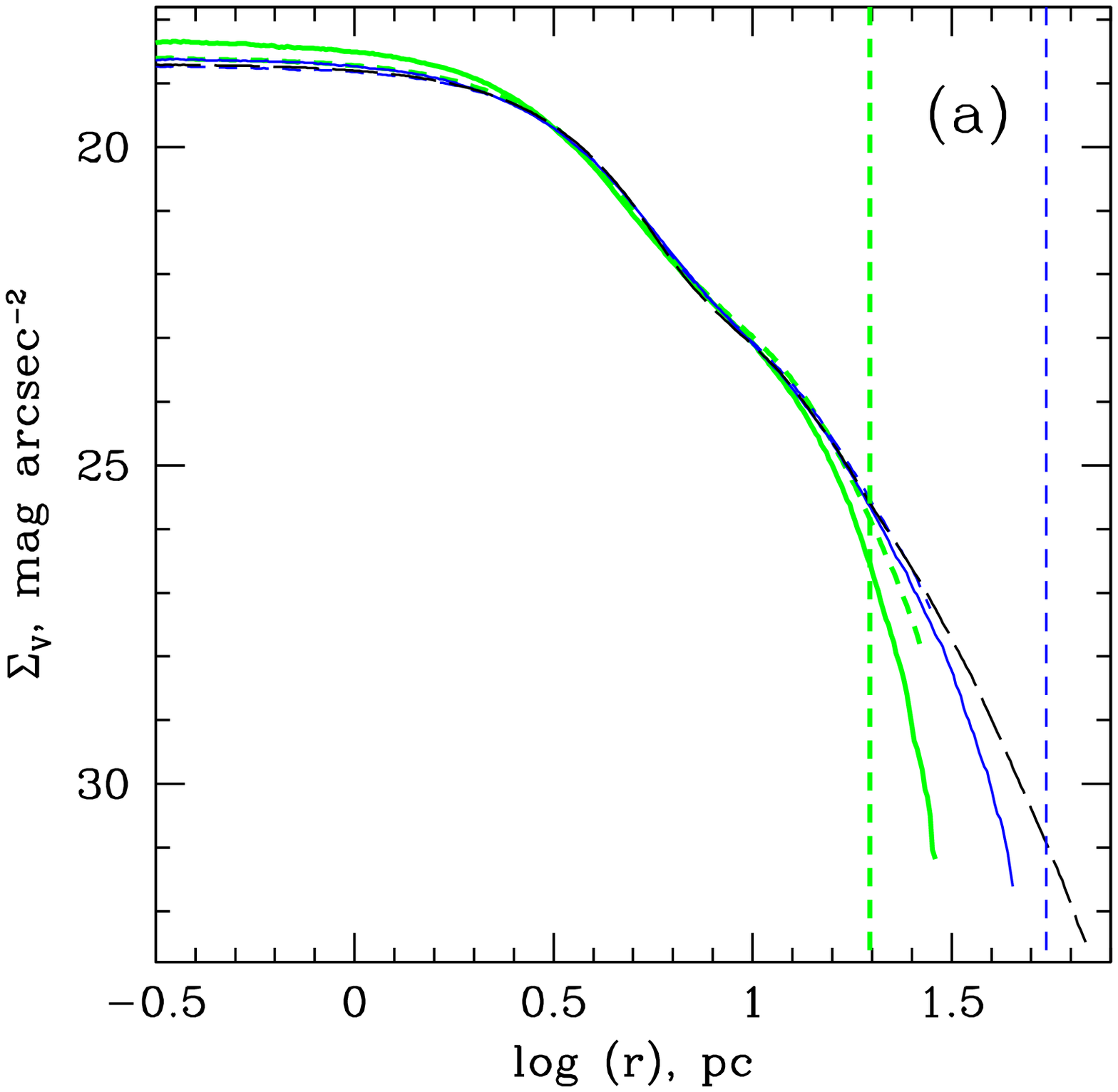}{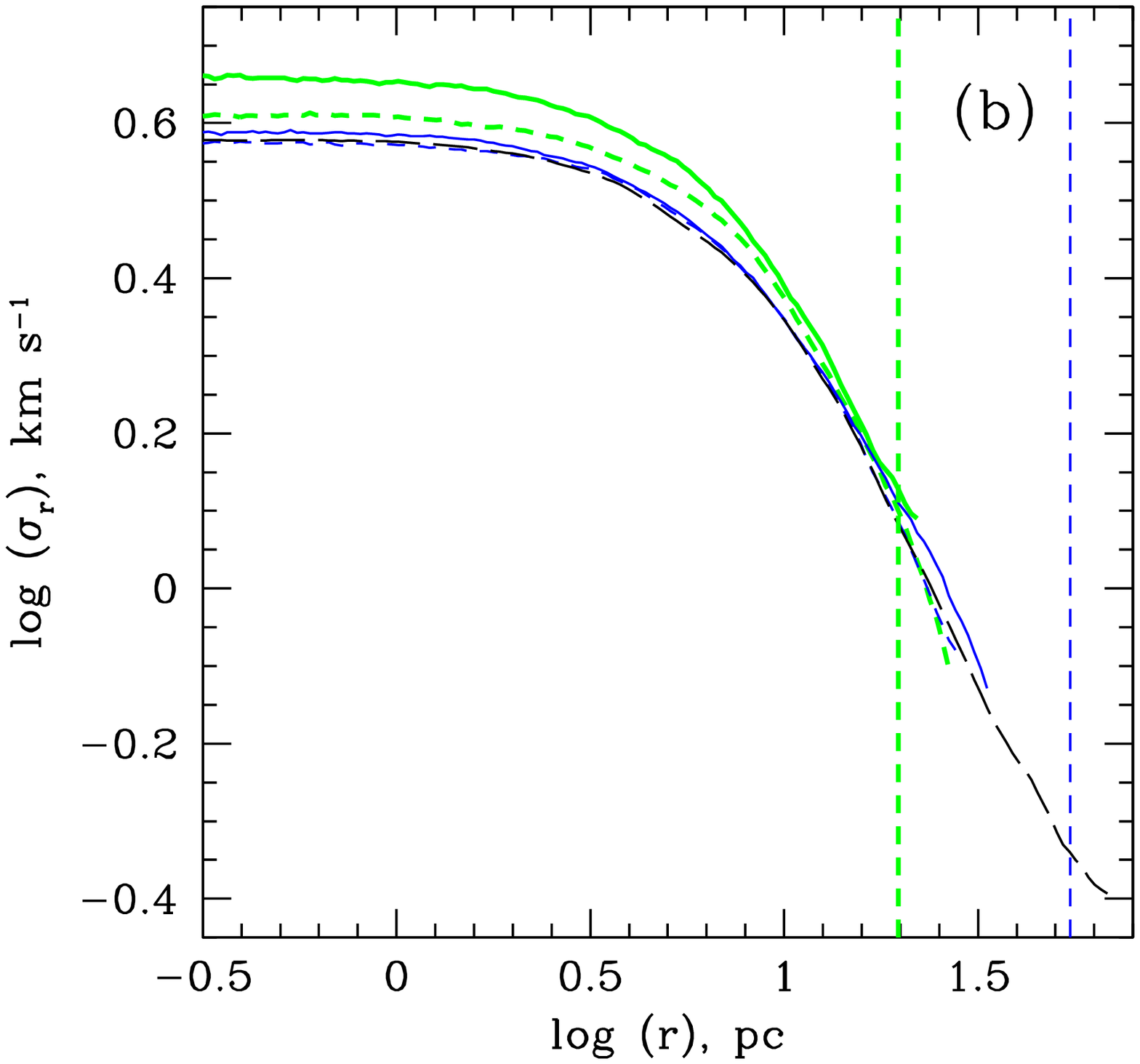}
\caption {
Radial profiles of observable quantities for warm collapse models. (a) Surface
brightness $\Sigma_V$. (b) Line-of-sight velocity dispersion $\sigma_r$.  Thick
lines (colored green in electronic edition) correspond to NFW cases; thin lines
(colored blue in electronic edition) correspond to Burkert cases.  Solid lines
show profiles for models W$_n$ and W$_b$, short-dashed lines correspond to
models W$_n$S and W$_b$S, and long-dashed lines correspond to model
W$_0$. Vertical short-dashed lines mark the values of $r_m$.
\label{proj_W} }
\end{figure*}


In Figure~\ref{rho_W} we show the radial density profiles for relaxed stellar
core and DM halo, both for NFW model (panel {\it a}) and Burkert model (panel
{\it b}), for the case of a warm collapse (``W'' models). Here long-dashed lines
show the density profiles for stars in the absence of DM (thin lines) and DM in
the absence of stars (thick lines). Solid lines, on the other hand, show the
density profiles for stars and DM having been evolved together.  We also show as
dotted lines the stellar radial density profiles for the case of a static DM
potential.

As you can see from Figure~\ref{rho_W}, in the absence of DM (model W$_0$), the
relaxed stellar cluster has a central flat core and a close to power-law outer
density profile. The radial density profile has also a small ``dent'' outside of
the core. It is not a transient feature (in the collisionless approximation), as
the virial ratio for the whole cluster is equal to unity within the measurements
errors ($\nu=1.001\pm 0.005$ for $t=t_1\dots t_2$) and the evolution time is
$>7$ crossing times even at the very last radial density profile point.  As we
show in Paper~II, collisional long-term dynamic evolution of a cluster will
remove this feature, bringing the overall appearance of the density profile
closer to that of the majority of observed GCs. The central stellar density in
the model W$_0$ is $\sim$ three orders of magnitude larger than the central
density of DM in the undisturbed Burkert halo, and becomes comparable to the
density of the undisturbed NFW halo only at very small radii $r\lesssim 0.1$~pc
(which is outside of the range of radial distances in Figure~\ref{rho_W}a).

In the case of a live DM halo coevolving with the stellar core, the DM density
profile is significantly modified within the central area dominated by stars
(see Figure~\ref{rho_W}). In both NFW and Burkert halos, DM is adiabatically
compressed by the potential of the collapsing stellar cluster to form a steeper
slope in the DM radial density profile. The slope of the innermost part of the
DM density profile is $\gamma\simeq -1.5$ for the NFW halo and $\gamma\simeq
-1.0$ for the Burkert halo.  The lack of resolution does not allow us to check
if the slope stays the same closer to the center of the DM halos.

In the outer parts of the DM halos (beyond the scale radius $r_s$, which is
marked by vertical dotted lines in Figure~\ref{rho_W}), the DM density
profile becomes steeper than the original slope of $\gamma=-3$. This can be
explained by the fact that our DM halo models are truncated at a finite radius $r_{\rm
vir}$ and hence are not in equilibrium. It should bear no effect on our results,
as in all our models stellar clusters do not extend beyond $r_s$.

As you can see in Figure~\ref{rho_W}, the impact of the presence of a live DM
halo on the stellar density profile is remarkably small (especially for the
Burkert halo). In the presence of DM, the central stellar density becomes
somewhat larger (by $\sim 60$\% and $\sim 10$\% for the NFW and Burkert halos,
respectively; see Table~\ref{tab3}). 

The most interesting effect is observed in the outer parts of the stellar
density profiles, where the slope of the profiles becomes significantly larger
than that of a purely stellar cluster, which is similar to the result of
\citet{pee84}. The density profile starts deviating from the profile for the
model W$_0$ somewhere between the radius $r_\rho$ and the radius $r_m$ (where
the enclosed masses for stars and DM become equal, see Table~\ref{tab3}).

There is a simple explanation for the steepening of the stellar density
profile seen in Figure~\ref{rho_W} around the radius $r_m$. In the absence of
DM, warm collapse of a homogeneous stellar sphere (our model W$_0$) is violent
enough to eject a number of stars into almost radial orbits, forming a halo of
the relaxed cluster. In the case when the DM halo is present, dynamics does not
change much near the center of the cluster where stars are the dominant mass
component. Around the radius $r_m$ the enclosed DM mass becomes comparable to
that of stars. As a result, gravitational deceleration, experienced by stars
ejected beyond $r_m$, more than doubles, resulting in a significantly smaller
number of stars populating the outer stellar halo at $r>r_m$. This should lead
to significantly steeper radial density profile beyond $r_m$, as observed in our
W$_{n,b}$ models.

As in \citet{pee84} model, in our model the steepening of the stellar density
profile is caused by the fact that at large radii the gravitational potential is
dominated by DM. The principal difference between the two models is that in
\citet{pee84} the stellar cluster is assumed to be isothermal and in equilibrium;
our clusters are not in equilibrium initially, and their final equilibrium
configuration is not isothermal (stars are dynamically colder in the outskirts
of the cluster).



To allow a direct comparison of the model results with observed
GCs, in Figure~\ref{proj_W} we plot both the surface brightness $\Sigma_v$
profile (panel {\it a}) and the line-of-sight velocity dispersion $\sigma_r$ profile
(panel {\it b}) for warm collapse models.  The surface brightness $\Sigma_V$ in
units of mag~arcsec$^{-2}$ is calculated using the following formula:

\begin{equation}
\Sigma_v = V_\odot - 2.5 [ \log (\zeta/\Upsilon_{\rm GC}) + 2 - 2 \log(3600\times 180/\pi)].
\label{eqSigma}
\end{equation}

\noindent Here $V_\odot=4\fm87$ is the absolute $V$-band magnitude of the Sun,
$\zeta$ is the projected surface mass density in units of $M_\odot$~pc$^{-2}$, and
$\Upsilon_{\rm GC}$ is the assumed $V$-band mass-to-light ratio for baryons in
GCs. For $\Upsilon_{\rm GC}$ we adopt the observationally derived
averaged value of 1.45 from \citet{mcl99}, which is also between the two
stellar-synthesis model values (1.36 for the Salpeter and 1.56 for the composite
--- with a zero slope for stellar masses $<0.3$~$M_\odot$ --- initial mass
functions) of \citet{mat98}.

As you can see in Figure~\ref{proj_W}, the presence of a DM halo introduces an
outer cutoff in the surface brightness radial profile of stellar cluster,
which makes the profile look very similar to a King model
profile. Interestingly, stellar cluster evolving in a live NFW DM halo acquires even
steeper outer density and brightness profiles than for the case of a static DM
potential (see Figures~\ref{rho_W}a and \ref{proj_W}a).

Analysis of the line-of-sight velocity dispersion profiles for warm collapse
models (Figure~\ref{proj_W}b) shows that the presence of a DM halo
slightly inflates the value of $\sigma_r$ in the core of the cluster (by
$\lesssim 20$\%, see Table~\ref{tab3}). The outer dispersion profile either
stays unchanged (as in the case of the Burkert halo), or becomes slightly
steeper (for the NFW model). The static DM models present profiles of an
intermediate type.

A slight increase of $\sigma_r$ in the central part of the cluster can
be misinterpreted observationally as a GC with no DM which has a bit larger
value of the baryonic mass-to-light ratio $\Upsilon$. To quantify
this effect, we apply a core fitting (or King's) method of finding the
central value of $\Upsilon$ in spherical stellar systems \citep{ric86}:

\begin{equation}
\Upsilon = \frac{9 \sigma_0^2}{2\pi G I_0 R_{hb}}.
\label{eqUpsilon}
\end{equation}

\noindent Here $I_0$ is the central surface brightness in physical units.
We assume here that $I_0=\zeta_0/\Upsilon_{\rm GC}$, where $\zeta_0$ is the
projected surface mass density at the center of the model cluster.  The core
fitting method assumes that the mass-to-light ratio is independent of radius,
which is obviously not true for our stars $+$ DM models.  In our models the
stellar particles have the same mass, so there is no radial mass segregation
between high and low mass stars caused by the long-term dynamic evolution of the
cluster. As a result, the equation~(\ref{eqUpsilon}) should not be used to
compare our models with real GCs. Instead, we use it to check if there is a
change in the apparent mass-to-light ratio $\Upsilon$ between our purely stellar
models and the models containing DM.  We list the values of $\Upsilon$ for
different models in Table~\ref{tab3}.  As you can see, in the case of NFW halo
(model W$_n$), the presence of the DM halo leads to $\sim 25$\% increase in the
value of the apparent mass-to-light ratio. Such a small increase is well within
the observed dispersion of $\Upsilon$ values for Galactic GCs \citep{pry93}. In
the case of the Burkert halo, the apparent central mass-to-light ratio is only
$\sim 6$\% larger than for the purely stellar model.

\subsection{Cold Collapse}


\begin{figure*}
\plottwo{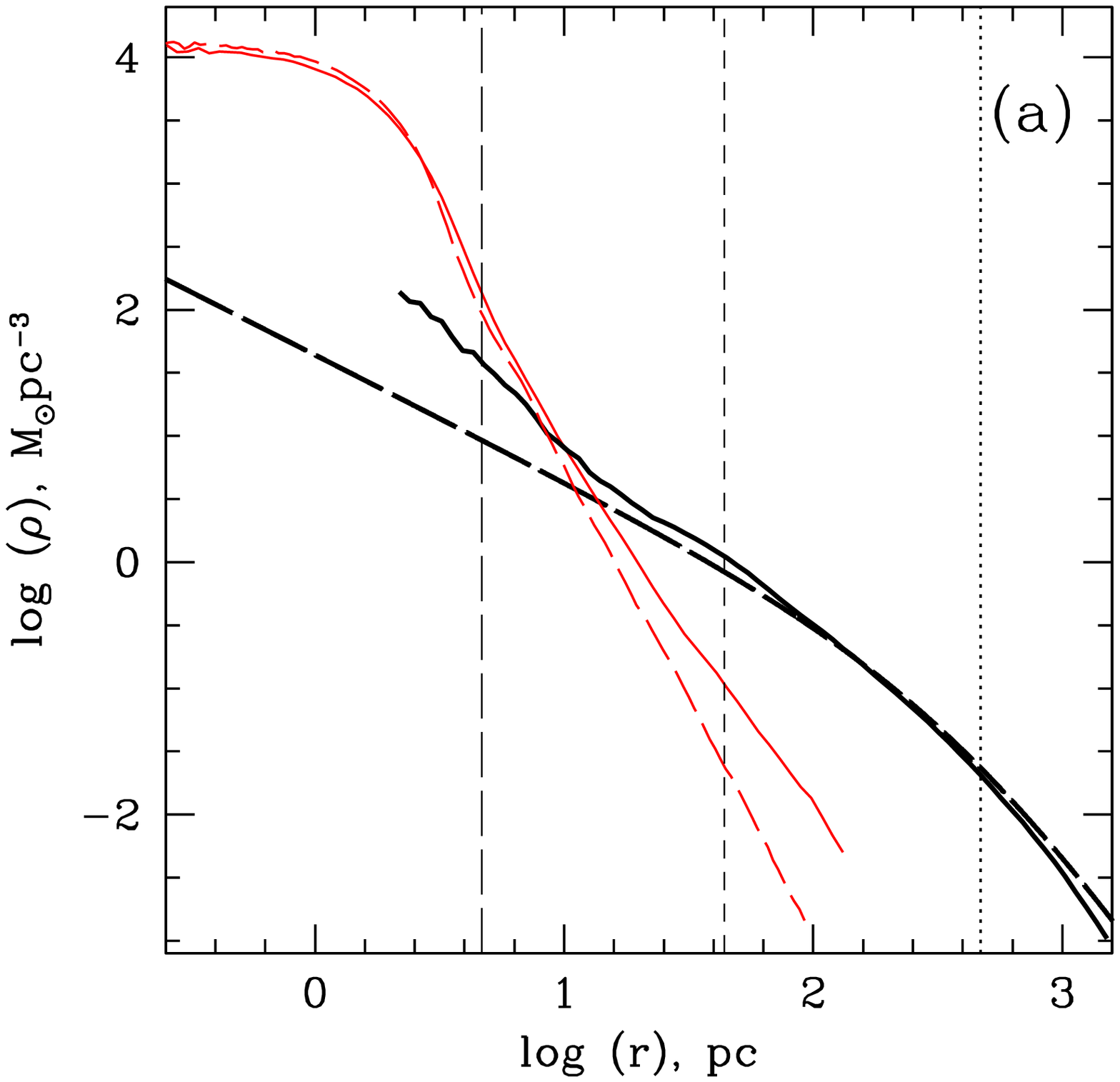}{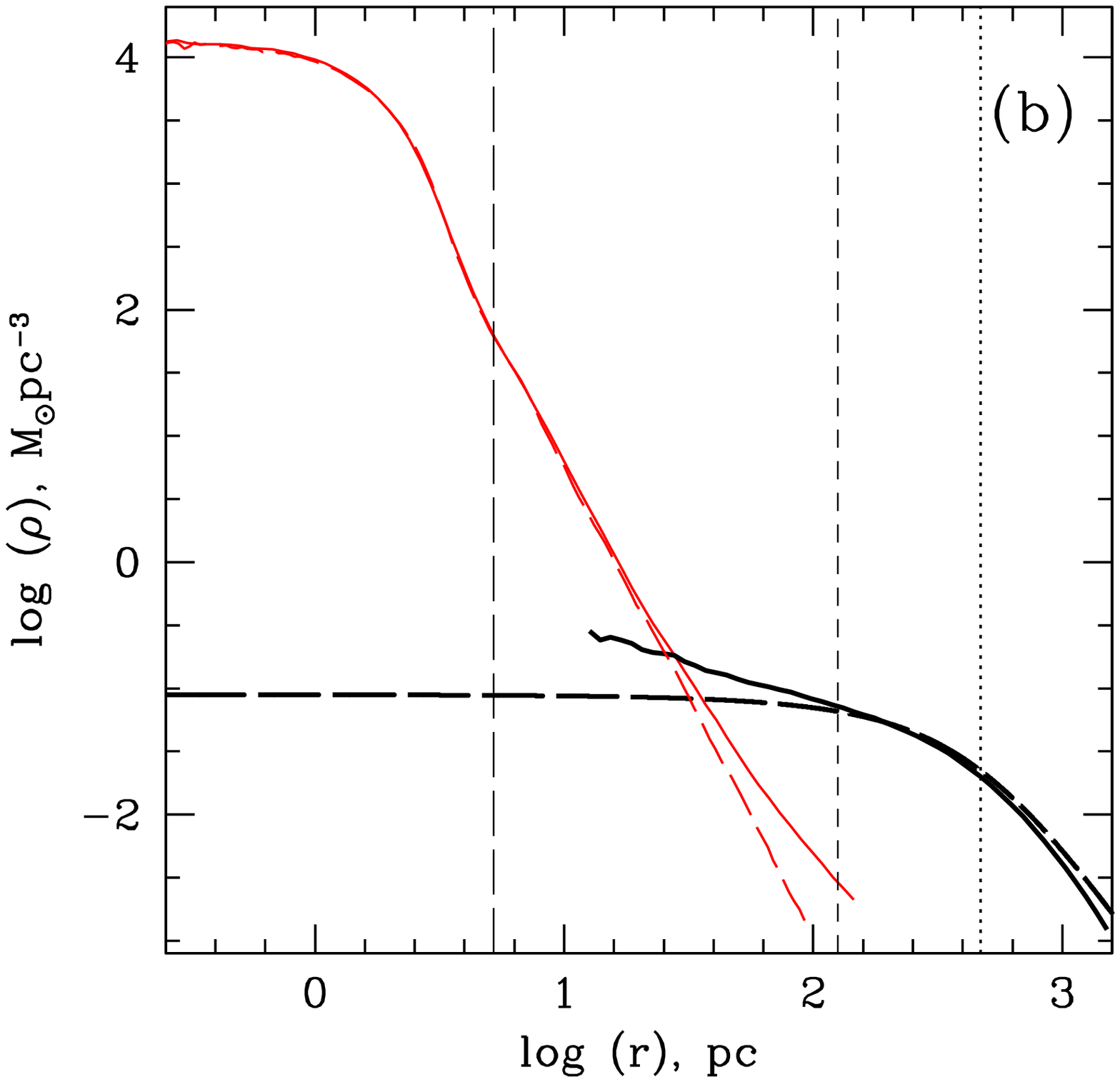}
\caption {
Radial density profiles for cold collapse models. (a) The case of NFW halo. (b)
The case of Burkert halo. Thick lines correspond to DM density; thin lines
(colored red in electronic edition) show stellar density. Solid lines correspond
to the cases of a live DM halo $+$ a stellar core (models C$_n$ and C$_b$);
long-dashed lines depict analytic DM profiles and a relaxed stellar cluster
profile in the absence of DM (model C$_0$).  Vertical long-dashed,
short-dashed, and dotted lines mark the values of $\epsilon_{\rm DM}$, $r_m$,
and $r_s$.
\label{rho_C} }
\end{figure*}

\begin{figure*}
\plottwo{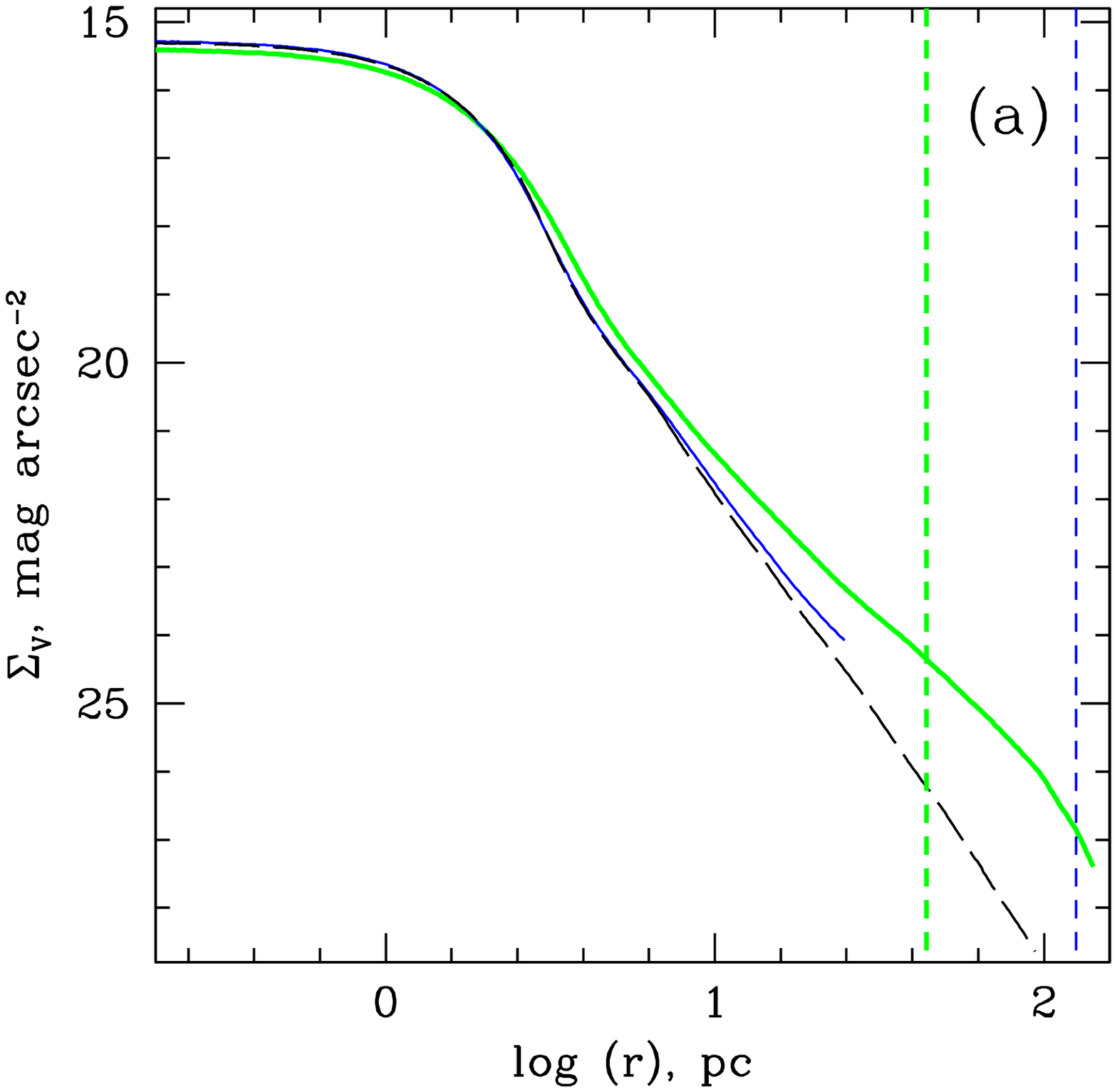}{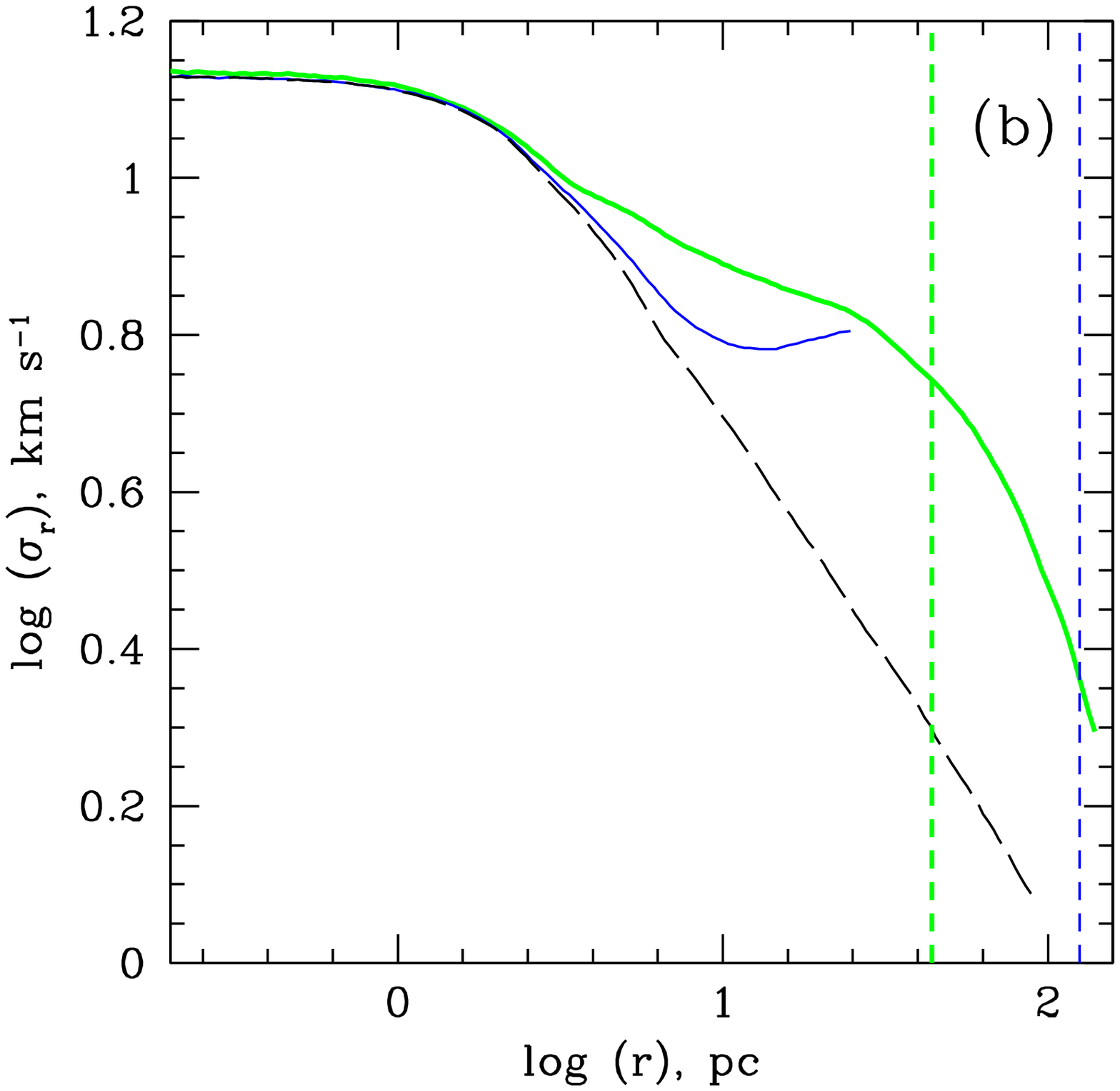}
\caption {
Radial profiles of observable quantities for cold collapse models. (a) Surface
brightness $\Sigma_V$. (b) Line-of-sight velocity dispersion $\sigma_r$.  Thick
lines (colored green in electronic edition) correspond to NFW cases; thin lines
(colored blue in electronic edition) correspond to Burkert cases.  Solid lines
show profiles for models C$_n$ and C$_b$; long-dashed lines correspond to model
C$_0$. Vertical short-dashed lines mark the values of $r_m$.
\label{proj_C} }
\end{figure*}


A principal difference between warm and cold collapse models from MS04 is
that in the cold collapse case a significant fraction of stars becomes unbound
after the initial relaxation phase (see Figure~\ref{beta_m}). In our model C$_0$
with the mass parameter $\beta=1.4$ the fraction of the stars lost is 30\%. The
radial density profile for C$_0$ model is similar to that of W$_0$ model (see
Figure~\ref{rho_C}). One can intuitively expect the escapers from the model
C$_0$ to be trapped inside our models containing DM, forming a distinctive
feature in the outer density and velocity dispersion profiles.

As you can see in Figures~\ref{rho_C} and
\ref{proj_C}, our cold collapse models do show such features. 
Both density and surface brightness profiles become more shallow in the outer
parts of the stellar clusters where DM dominates stars. This is valid for both NFW and
Burkert halos. For the NFW and Burkert cases, the slope of the outer stellar density
profile is $\gamma=-2.6$ and $-2.2$, respectively. For the purely stellar case
(model C$_0$) the slope is $\gamma=-3.8$, so the relative change in the slope is
$\Delta \gamma=1.2$ and $1.6$ for the NFW and Burkert cases. This behavior is
mimicked by the corresponding surface brightness profiles
(Figure~\ref{proj_C}a). It is interesting to note that, for the C$_n$ model the
radial $\Sigma_V$ profile exhibits a significant steepening of the slope in the
outmost parts of the cluster, creating an appearance of a ``tidal cutoff''
(similarly to the warm collapse cases).

Even more pronounced are features in the line-of-sight velocity
dispersion profiles (Figure~\ref{proj_C}b). In both NFW and Burkert
cases, there is a plateau in the radial $\sigma_r$ profiles around the radius
$r_m$ where DM becomes the dominant mass component. The apparent
``break'' in the surface brightness profile and accompanying flattening
of the radial $\sigma_r$ profile, seen in our models containing DM,
can be misinterpreted observationally as a presence of ``extratidal'' stars
heated by the tidal field of the host galaxy. 

Similarly to warm collapse models, the apparent mass-to-light ratio $\Upsilon$
for cold collapse models in the presence of a live DM halo is very close to the
purely stellar case C$_0$ (see Table~\ref{tab3}). Interestingly, the half-mass
radii for warm and cold collapse models with DM are almost identical. This is
consistent with the properties of the observed GCs, which have comparable
half-mass radii for a wide range of cluster masses.

DM density profiles for the cold collapse cases (Figure~\ref{rho_C}) exhibit a
very similar behavior to the warm collapse models in the stars dominated central
region. For both NFW and Burkert halos, the innermost slopes of the density
profiles become steeper ($\gamma\simeq -1.8$ and $-0.6$, respectively) in the
presence of a stellar core. The NFW DM density profile shows a break around the
radius $r_m$.

As you can see in Table~\ref{tab3} and Figure~\ref{rho_C}, the final stellar
cluster half-mass radius is smaller than the DM softening length $\epsilon_{\rm
DM}\sim 5$~pc.  We reran models C$_n$ and C$_b$ with much smaller value of
$\epsilon_{\rm DM}=1$~pc to test the possibility that our results were
influenced by the fact that DM is not resolved on the stellar cluster scale.
For both C$_n$ and C$_b$, the relaxed stellar profiles are found to be virtually
identical to the cases with larger $\epsilon_{\rm DM}$. In particular, a
``kink'' seen in Figures~\ref{rho_C}a and \ref{rho_C}b around the radius
$\epsilon_{\rm DM}$ is also present at the same location in the simulations
with much smaller value of $\epsilon_{\rm DM}$, and is definitely not a
numerical artifact.  In the case of NFW halo, the central stellar velocity
dispersion becomes slightly larger, which results in somewhat larger value of
$\Upsilon=1.88$. This could be in part because of artificial mass segregation,
which should be more pronounced in the case of $\epsilon_{\rm DM}$ being
significantly lower than the optimal value (in our cold collapse models, DM
particles are $\sim 20$ times more massive than stellar particles). As a result,
the actual $\Upsilon$ value could be even closer to the baryonic value. In the
case of Burkert halo with $\epsilon_{\rm DM}=1$~pc, the core mass-to-light ratio
$\Upsilon=1.45$, which is identical to the case of no DM. We conclude that our
choice of $\epsilon_{\rm DM}$ did not affect the main results presented in this
Section.

\subsection{Hot Collapse}

It is generally assumed that in an equilibrium star-forming cloud no bound
stellar cluster will be formed after stellar winds and supernova explosions
expel the remaining gas if the star formation efficiency is less than 50\%
(which corresponds to the initial virial parameter for the stellar cluster of
$\nu>2$).  In MS04 we showed that it is not true for initially homogeneous
stellar clusters which have a Maxwellian distribution of stellar velocities. In
such clusters with the initial virial parameter as large as 2.9 (corresponding
to the mass parameter $\beta$ as low as $-0.7$), a bound cluster, containing
less than 100\% of the total mass, is formed by the slowest moving stars.  (A
similar conclusion was reached by \citealt{boi03} for initially polytropic
stellar spheres.)


\begin{figure*}
\plottwo{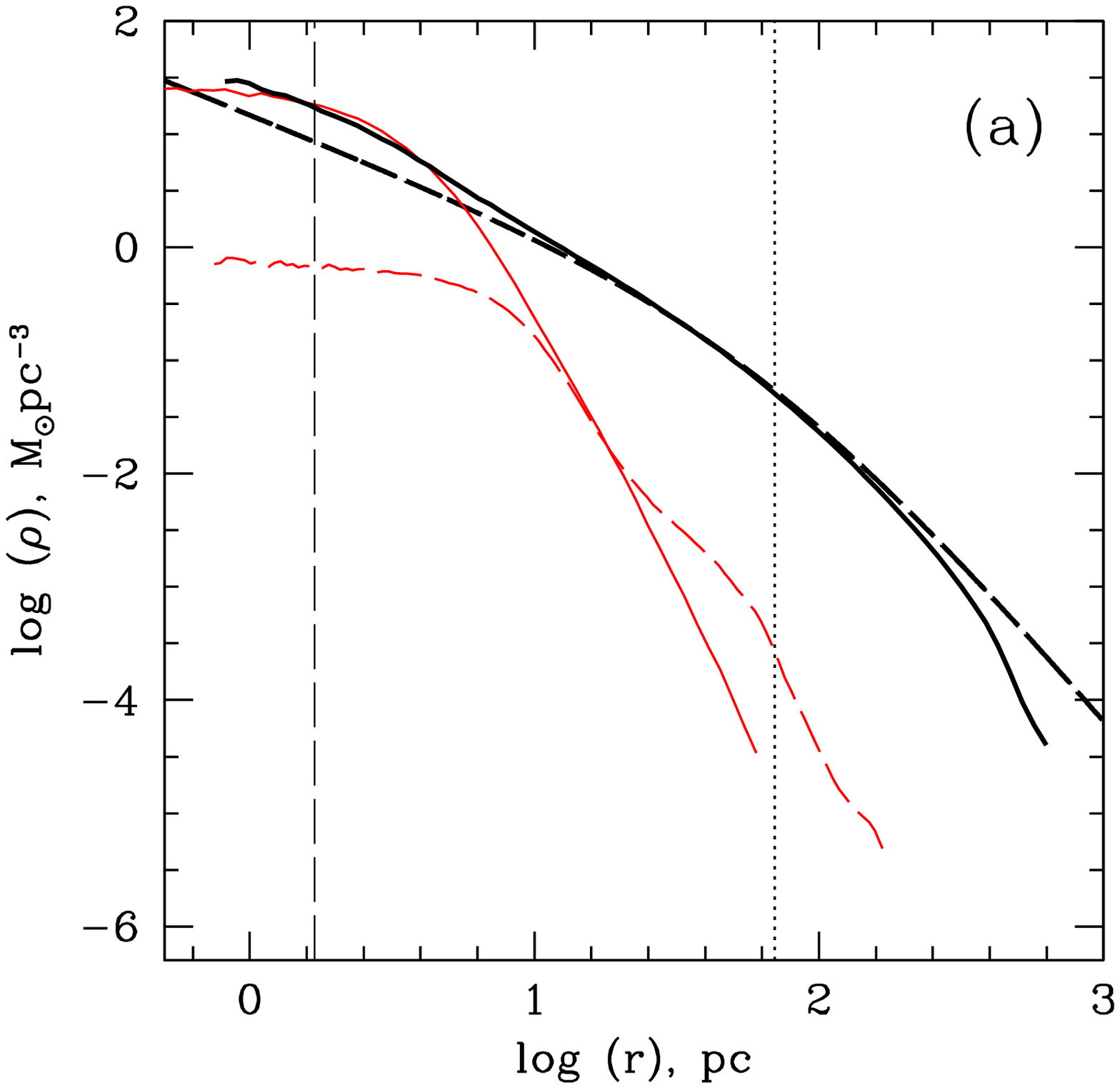}{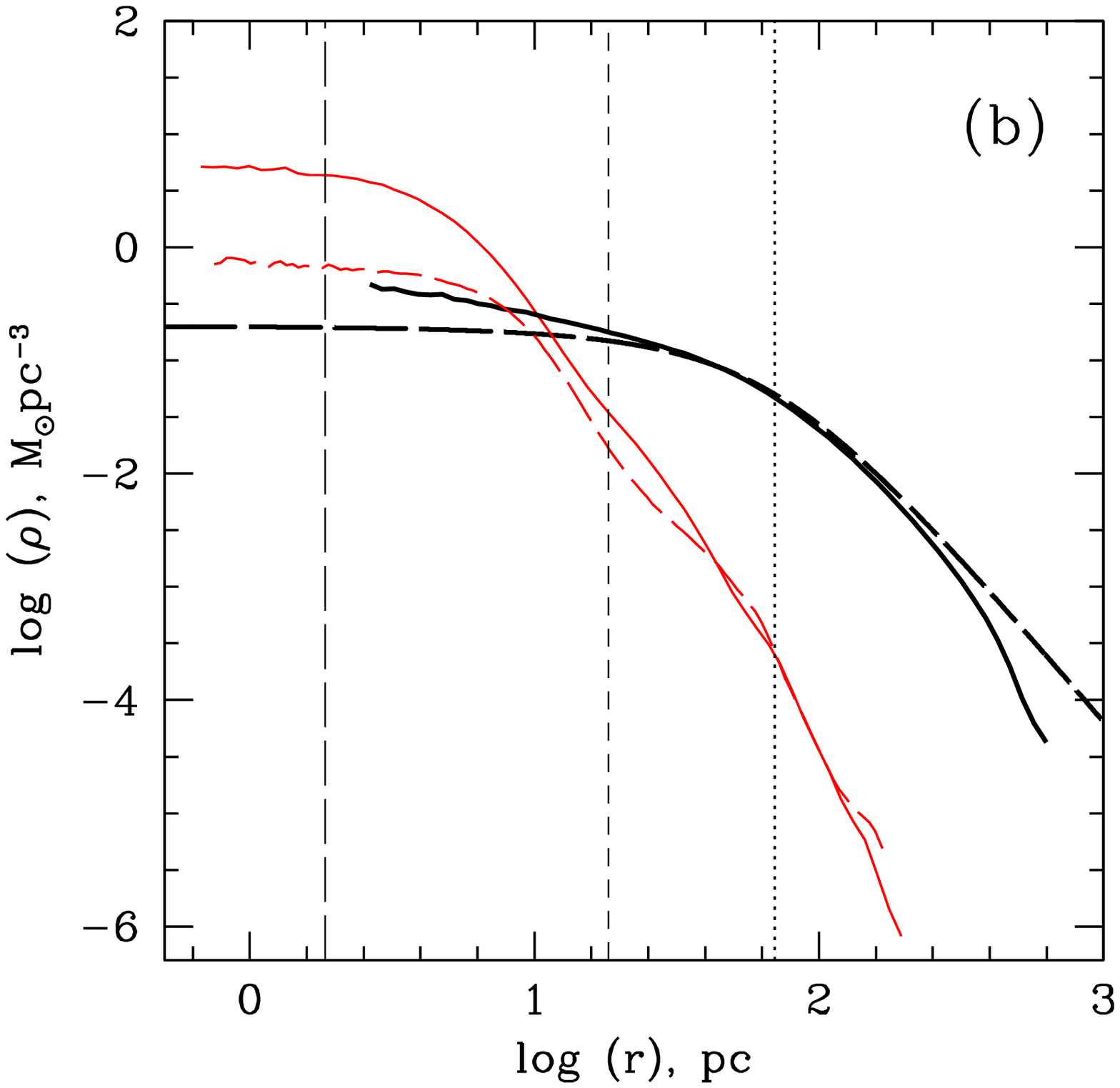}
\caption {
Radial density profiles for hot collapse models. (a) The case of NFW halo. (b)
The case of Burkert halo. Thick lines correspond to DM density; thin lines
(colored red in electronic edition) show stellar density. Solid lines correspond
to the cases of a live DM halo $+$ a stellar core (models H$_n$ and H$_b$);
long-dashed lines depict analytic DM profiles and a relaxed stellar cluster
profile in the absence of DM (model H$_0$). Vertical long-dashed, short-dashed,
and dotted lines mark the values of $\epsilon_{\rm DM}$, $r_m$ (only the Burkert
case is shown, as in the NFW case the enclosed DM mass is larger than that of
stars at any radius), and $r_s$.
\label{rho_H} }
\end{figure*}

\begin{figure*}
\plottwo{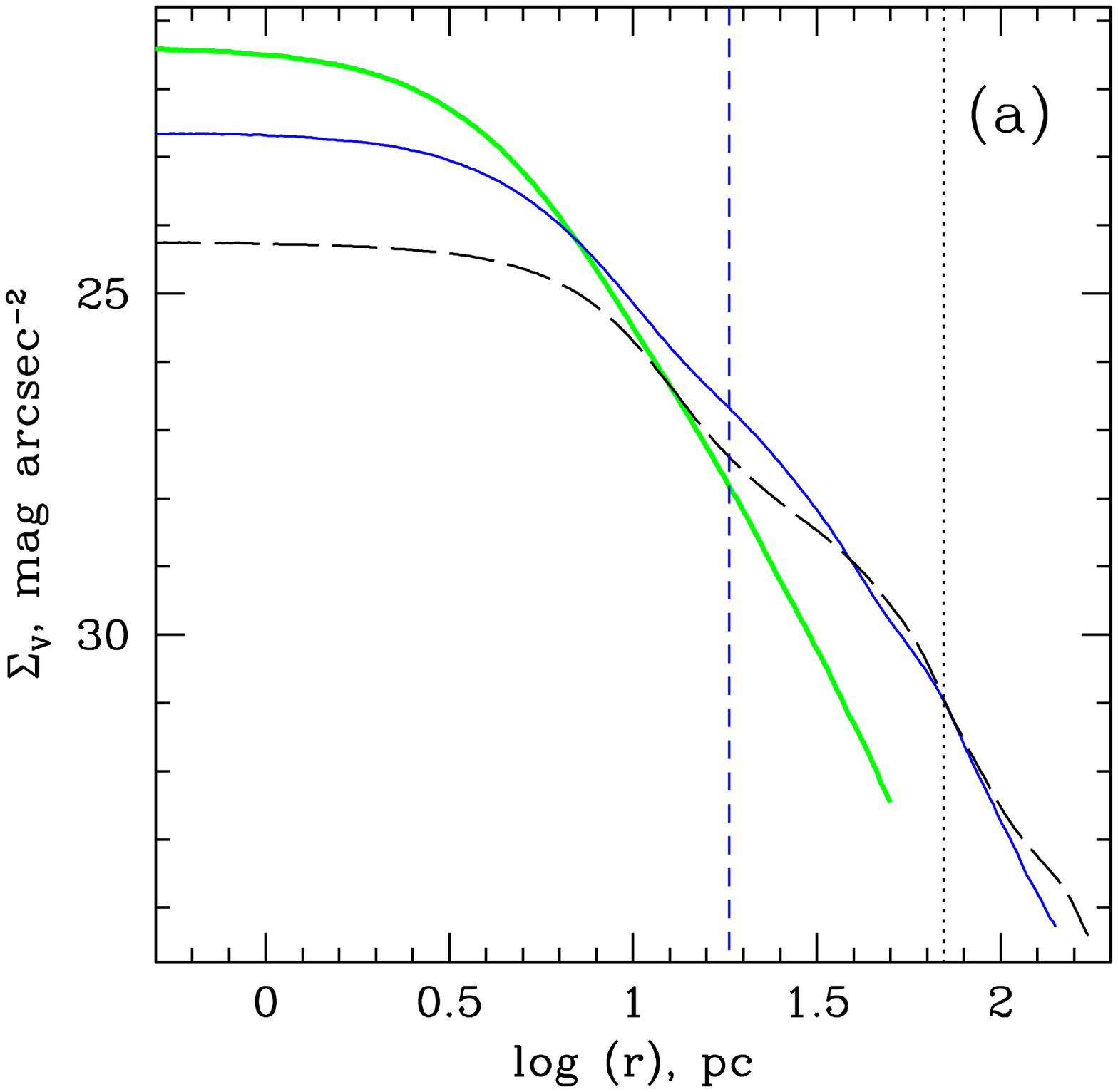}{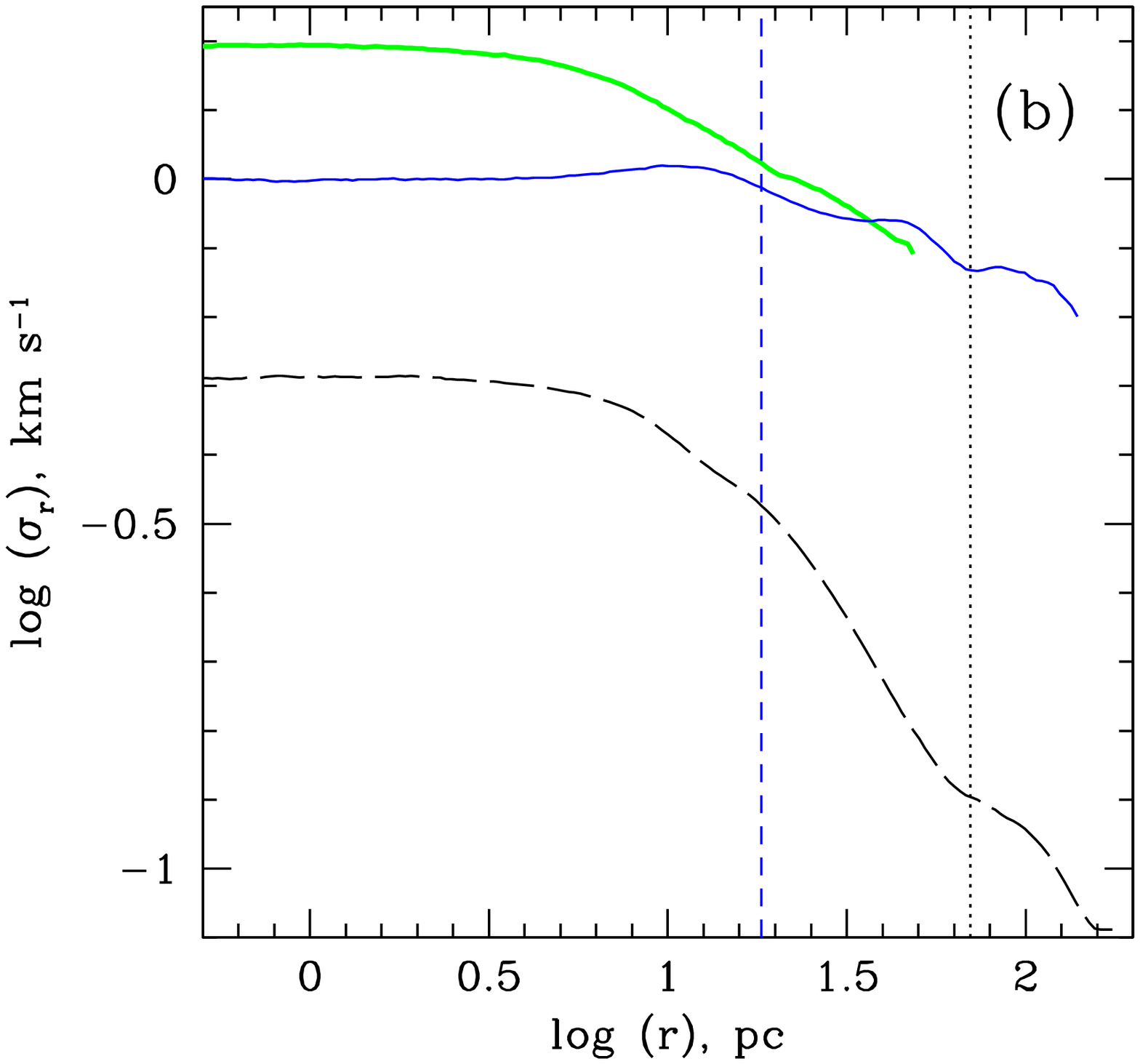}
\caption {
Radial profiles of observable quantities for hot collapse models. (a) Surface
brightness $\Sigma_V$. (b) Line-of-sight velocity dispersion $\sigma_r$.  Thick
lines (colored green in electronic edition) correspond to NFW cases; thin lines
(colored blue in electronic edition) correspond to Burkert cases.  Solid lines
show profiles for models H$_n$ and H$_b$; long-dashed lines correspond to model
H$_0$. Vertical short-dashed line marks the value of $r_m$ (only the Burkert
case is shown, as in the NFW case the enclosed DM mass is larger than that of
stars at any radius); vertical dotted line corresponds to $r_s$.
\label{proj_H} }
\end{figure*}


In our model H$_0$ ($\beta=-0.6$, $\nu=2.5$), $\sim 57$\% of all stars stay
gravitationally bound after the initial relaxation phase (see
Figure~\ref{beta_m}). Because in both cold and hot collapse models for purely
stellar clusters a significant fraction of stars become unbound, one might
expect that stellar density and velocity dispersion profiles to be similar for
these two cases in the presence of DM. Analysis of Figures~\ref{rho_H} and
\ref{proj_H} shows that it is definitely not the case: H$_{n,b}$ models show
completely different behavior from C$_{n,b}$ models. The explanation is simple:
C$_0$ model starts with the stellar density which is already larger than the
density of DM within the stellar core in models C$_{n,b}$ (in other words,
$S>1$, see Table~\ref{tab1}), and after the relaxation phase reaches even higher
density which is $\sim (r_*/r_{h,*})^3/2\simeq 430$ times higher than the initial
density. The hot collapse model H$_0$, on the other hand, having started with
$S>1$, acquires a density which is $\sim 120$ times lower than the initial
density after the expansion and relaxation phase, resulting in $S<1$ for the
relaxed stellar cluster even for the Burkert halo. Under these circumstances,
radial profiles for H$_n$ and H$_b$ models should be significantly different
from the case with no DM (H$_0$), which is what is seen in Figures~\ref{rho_H} and
\ref{proj_H}. The only common feature between the models C$_{n,b}$ and
H$_{n,b}$ is that in both cases all would-be escapers are trapped by the
potential of the DM halo.

In the case of the NFW halo, central stellar density and central velocity
dispersion become larger in the presence of DM by factors of 35 and 3,
respectively. In the Burkert halo case, the increase is not as dramatic, but
still significant: central density and velocity dispersion become 7 and 2 times
larger than those in the absence of DM. Interestingly, as in the cold collapse cases,
the presence of DM brings the half-mass radius of the stellar core in H$_{n,b}$
models closer to the half-mass radius of the warm collapse models (see
Table~\ref{tab3}). The apparent central mass-to-light ratio $\Upsilon$ is
noticeably inflated by the presence of DM (especially for the case of NFW halo)
--- by 76\% and 36\% for NFW and Burkert models, respectively.

As you can see in Figure~\ref{rho_H}a, in the model H$_n$ stars have
a comparable density to DM in the core, and become less dense than DM outside of
the core radius $r_0$. The situation is different (and more in line with warm
and cold collapse models) for the H$_b$ model (Figure~\ref{rho_H}b),
where stars are the dominant (though not by a large margin) mass component in
the core.  There are no obvious features in the density and surface brightness
profiles caused by the presence of DM: the profiles look very similar to W$_0$
and C$_0$ models which do not have DM.

Similarly to warm and cold collapse cases, the presence of a stellar cluster
makes DM denser in the central area. In the case of NFW halo, the innermost
slope of the DM density profile is comparable to the slope of $\gamma=-1$ for
undisturbed halo. For the Burkert halo, the slope becomes steeper, reaching the
value of $\gamma\sim -0.4$ at the innermost resolved point.

The most interesting behavior is exhibited by the line-of-sight velocity
dispersion profiles (Figure~\ref{proj_H}b). In the presence of DM,
the profiles become remarkably flat, changing by mere 0.2 -- 0.3~dex over the
whole range of radial distances (out to $\sim 15$ apparent half-mass radii).
The combination of the low mass, low central surface brightness, large
apparent mass-to-light ratio, and almost flat velocity dispersion
profiles seen in models H$_{n,b}$ can be mistakingly interpreted as a
GC observed at the final stage of its disruption by the tidal forces of
the host galaxy.

\section{DISCUSSION AND CONCLUSIONS}
\label{discussion}

A significant hindrance to a wider acceptance of the primordial scenarios for
GC origin is an apparent absence of DM in Galactic GCs.  Many observational
facts have been suggested to be evidence for GCs having no DM, including the
presence of such features in the outer parts of the GC density profiles as
apparent tidal cutoffs or breaks, relatively low values of the apparent
central mass-to-light ratio $\Upsilon$ which are consistent with purely baryonic
clusters, flat radial distribution of the line-of-sight velocity dispersion in
the outskirts of GCs believed to be a sign of tidal heating, and non-spherical
shape of the clusters in their outer parts.

Here we present the results of simulations of stellar clusters relaxing inside
live DM minihalos in the early universe (at $z=7$). We study three distinctly
different cases which can correspond to very different gas-dynamic processes
forming a GC: a mild warm collapse, a violent cold collapse resulting in a much
denser cluster with a significant fraction of stars escaping the GC in the
absence of DM, and a hot collapse resulting in a lower density cluster with many
would-be stars-escapers. We show that GCs forming in DM minihalos exhibit the
same properties as one would expect from the action of the tidal field of the
host galaxy on a purely stellar cluster: King-like radial density cutoffs (for
the case of a warm collapse), and breaks in the outer parts of the density
profile accompanied by a plateau in the velocity dispersion profile (for a cold
collapse).  Also, the apparent mass-to-light ratio for our clusters with DM is
generally close to the case of a purely stellar cluster. (The special case of a
hot collapse inside a DM halo which produces inflated values of $\Upsilon$ can
be mistaken for a cluster being at its last stage of disruption by the tidal
forces.)

We argue that increasingly eccentric isodensity contours observed in the
outskirts of some GCs could be created by a stellar cluster relaxing inside a
triaxial DM minihalo, and not by external tidal fields as it is usually
interpreted. Indeed, cosmological DM halos are known to have noticeably
non-spherical shapes; a stellar cluster relaxing inside such a halo would have
close to spherical distribution in its denser part where the stars dominate DM,
and would exhibit isodensity contours of increasingly larger eccentricity in its
outskirts where DM becomes the dominant mass component.

It is also important to remember that few Galactic GCs show clear signs of a
``tidal'' cutoff in the outer density profiles \citep	{TKD95}. The ``tidal''
features of an opposite nature --- ``breaks'' in the outer parts of the radial
surface brightness profiles in some GCs --- are often observed at or below the
inferred level of contamination by foreground/background objects, and could be
in many cases an artifact of the background subtraction procedure which relies
heavily on the assumption that the background objects are smoothly distributed
across the field of view. A good example is that of Draco dwarf spheroidal
galaxy. \citet{irw95} used simple non-filtered stellar counts from photographic
plates followed by a background subtraction procedure to show that this galaxy
appears to have a relatively small value of its radial density ``tidal cutoff''
$r_t=28.3\pm 2.4$~arcmin and a substantial population of ``extratidal'' stars.
\citet{ode01} used a more advanced approach
of multi-color filtering of Draco stars from the Sloan Digital Sky Survey images
and achieved much higher signal-to-noise ratio than in \citet{irw95}. New,
higher quality results were supposed to make the ``extratidal'' features of
Draco much more visible. Instead, \citet{ode01} demonstrated that the Draco's
radial surface brightness profile is very regular down to a very low level (0.003
of the central surface brightness), and suggested a larger value for the King
tidal radius of $r_t\simeq 50$~arcmin.

We argue that the qualitative results presented in this paper are very general,
and do not depend much on the the fact that we used MS04 model to set up the
initial non-equilibrium stellar core configurations, and on the particular values
of the model parameters (such as $\rho_{i,*}$, $\sigma_{i,*}$, and $\chi$). As
we discussed in Section~\ref{results}, the appearance of ``tidal'' or
``extratidal'' features in our warm and cold collapse models is caused by two
reasons: (1) at radii $r>r_m$ the potential is dominated by DM, whereas in the
stellar core the potential is dominated by stars from the beginning till the end
of simulations, and (2) the collapse is violent enough to eject a fraction of
stars beyond the initial stellar cluster radius. 

As we showed in \S~\ref{physical}, for the initial stellar density
$\rho_{i,*}=14$~$M_\odot$~pc$^{-3}$ (from MS04), the whole physically plausible
range of stars-to-DM mass ratios $\chi$ and the GC formation redshifts $z$
satisfy the above first condition. For warm and cold collapse, this condition
can be reexpressed as $S\equiv m_*/m_{\rm DM}[r_*]>1$. One can easily estimate
$S$ for other values of $\rho_{i,*}$.

The second condition is more difficult to quantify. In MS04 we showed that
collapsing homogeneous isothermal spheres produce extended halos for any values
of the initial virial ratio $\nu$ (except for $\nu>2.9$ systems which are too
hot to form a bound cluster in the absence of DM). \citet{roy04} simulated cold
collapse for a wider spectrum of initial cluster configurations, including
power-law density profile, clumpy and rotating clusters. In all their
simulations an extended halo is formed after the initial violent relaxation.
It appears that in many (probably most) stellar cluster configurations, which are
not in detailed equilibrium initially, our second condition can be met.

Of course, not every stellar cluster configuration will result in a GC-like
object, with a relatively large core and an extended halo, after the initial
violent relaxation phase. \citet{spi72} demonstrated that a warm ($\nu=0.5$)
collapse of a homogeneous isothermal sphere produces clusters with large cores
(their models D and G). Adiabatic collapse of homogeneous isothermal spheres
produces clusters which surface density profiles are very close to those of
dynamically young Galactic GCs (MS04). \citet{roy04} concluded that any cold
stellar system which does not contain significant inhomogeneities relaxes to a
large-core configuration. The results of \citet{roy04} can also be used to
estimate the importance of adiabaticity for core formation. Indeed, their
initially homogeneous models H and G span a large range of $\nu$, and include
both adiabatic cases ($\nu>0.16$ for their number of particles $N=3\times 10^4$,
from eq.[\ref{eqbeta}]) and non-adiabatic ones ($\nu<0.16$). It appears that all
their models (adiabatic and non-adiabatic) form a relatively large core. The
issue is still open, but it appears that the adiabaticity requirement (our
eq.[\ref{eqbeta}]) is not a very important one for our problem. (Though this
requirement is met automatically for real GCs for the values of $\rho_{i,*}$ and
$\sigma_{i,*}$ derived in MS04.)

An important point to make is that the simulations presented in this paper
describe the collisionless phase of GC formation and evolution, and cannot be
directly applied to GCs which have experienced significant secular evolution due
to encounters between individual stars. In MS04 we showed that such
collisionless simulations of purely stellar clusters describe very well the
surface brightness profiles of dynamically young Galactic GCs (such as
NGC~2419, NGC~5139, IC~4499, Arp~2, and Palomar~3 -- see Fig.~1 in MS04).  In
Paper~II we will address (among other things) the issue of long-term dynamic
evolution of hybrid GCs. We will demonstrate that, at least for the
warm-collapse case, secular evolution does not change our qualitative
results presented in this paper.

In the light of the results presented in this paper and the above arguments, we
argue that additional observational evidence is required to determine with any
degree of confidence if GCs have any DM presently attached to them, or if they
are purely stellar systems truncated by the tidal field of the host galaxy. A
decisive evidence would be the presence of obvious tidal tails. A beautiful
example is given by Palomar~5 \citep{ode03}, where tidal tails were observed to
extend over $10\degr$ in the sky.

Even if a GC cluster is proven not to have a significant amount of DM, it does
not preclude it having been formed originally inside a DM minihalo. In
``semi-consistent'' simulations of \citet{bro02} of a dwarf galaxy formation,
proto-GCs were observed to form inside DM minihalos, with the DM being lost
during the violent relaxation accompanying the formation of the dwarf galaxy.
Unfortunately, their simulations did not have enough resolution to
clarify the fate of DM in proto-GCs. In Paper~II we will address the issue of
the fate of DM in hybrid proto-GCs experiencing severe tidal stripping in the
potential of the host dwarf galaxy.

\acknowledgements

\begin{appendix}

\section{Distribution function for Burkert halos}
\label{app1}

The dimensionless potential $\Psi$ of Burkert halos with the density profile given by
equations~(\ref{rho_b})--(\ref{rho0_b}) is as follows:

\begin{equation}
\label{psibur}
\Psi = 1 - \frac{2}{\pi} \left\{(1+R^{-1})\left[\arctan R - \ln (1+R)\right] + 
\frac12 (1-R^{-1})\ln (1+R^2)\right\}.
\end{equation}

\noindent Here $R\equiv r/r_s$ is the radial distance in scale radius units, and
$\Psi$ is in $\pi^2 G \rho_{0,b} r_s^2$ units. The potential $\Psi$ is equal
to 1 at the center of the halo, and zero in the infinity.

Phase-space distribution function for a Burkert halo with an isotropic
dispersion tensor can be derived through an Abel transform
\citep[p. 237; we skip the second part of the integrant, which is equal to 
zero for Burkert halos]{bin87}:

\begin{equation}
\label{FE}
F(E)=\frac{1}{\sqrt{8}\pi^2}\int_0^E
\frac{d^2\rho}{d\Psi^2}\frac{d\Psi}{(E-\Psi)^{1/2}}.
\end{equation}

\noindent Here $E$ is the relative energy in the same units as $\Psi$ and $\rho$ is the
density in $\rho_{0,b}$ units, which results in $(\pi
r_s)^3\rho_{0,b}^{1/2}G^{3/2}$ units for $F$.

The function $d^2\rho/d\Psi^2$ for a Burkert halo has the same asymptotic
behavior for $R\rightarrow 0$ and $R\rightarrow \infty$ as the model~I of
\citet{wid00}, which has the density profile $\rho\propto (1+R)^{-3}$:

\begin{equation}
\frac{d^2\rho}{d\Psi^2}(R\rightarrow 0) \propto R^{-3}; \quad
\frac{d^2\rho}{d\Psi^2}(R\rightarrow \infty) \propto \frac{\Psi}{(-\ln \Psi)^3}.
\end{equation}

\noindent This allowed us to use the following analytic fitting formula of \citet{wid00}
for the distribution function of Burkert halo:

\begin{equation}
\label{wid}
F(E)=F_0 E^{3/2}(1-E)^{-1}\left(\frac{-\ln E}{1-E}\right)^q e^P.
\end{equation}

\noindent Here $P\equiv \sum_i p_i E^i$ is a polynomial introduced to improve the fit.

We solve equation~(\ref{FE}) numerically for the interval of radial
distances $R=10^{-6}\dots 10^6$. We obtained the following values for the
fitting coefficients in equation~(\ref{wid}):

\begin{eqnarray}
F_0=5.93859\times 10^{-4}; \quad & q=-2.58496; \quad & p_1=-0.875182;\nonumber\\
p_2=24.4945;   \quad & p_3=-147.871; \quad & p_4=460.351;\label{fit}\\
p_5=-747.347; \quad & p_6=605.212; \quad & p_7=-193.621.\nonumber
\end{eqnarray}

We checked the accuracy of the fitting formula~(\ref{wid}) with the above values
of the fitting coefficients by comparing the analytic density profile of the
Burkert halo (eq.~[\ref{rho_b}]) with the density profile derived from the
distribution function
\citep[p. 236]{bin87}:

\begin{equation}
\rho(R)=2\sqrt{8}\pi^2\int_0^\Psi F(E) (\Psi-E)^{1/2}dE.
\label{rho_R}
\end{equation}

\noindent The deviation of $\rho(R)$ in equation~(\ref{rho_R}), with $F(E)$ given 
by equations~(\ref{wid})--(\ref{fit}), from the analytic density profile was
found to be less than 0.7\% for the interval of the radial distances
$R=10^{-6}\dots 10^6$.

\section{Projection method}
\label{app2}

For $N$-body models of spherically symmetric stellar systems, such as GC models
presented in the paper, one can produce radial profiles of projected observable
quantities (such as surface mass density or velocity dispersion) by averaging
these quantities over all possible rotations of the cluster relative to the
observer. Radial profiles created in this way preserve maximum information,
which is very important for studies of the outskirts of the clusters and for
clusters simulated with relatively small number of particles.


\begin{figure}
\epsscale{0.4}
\plotone{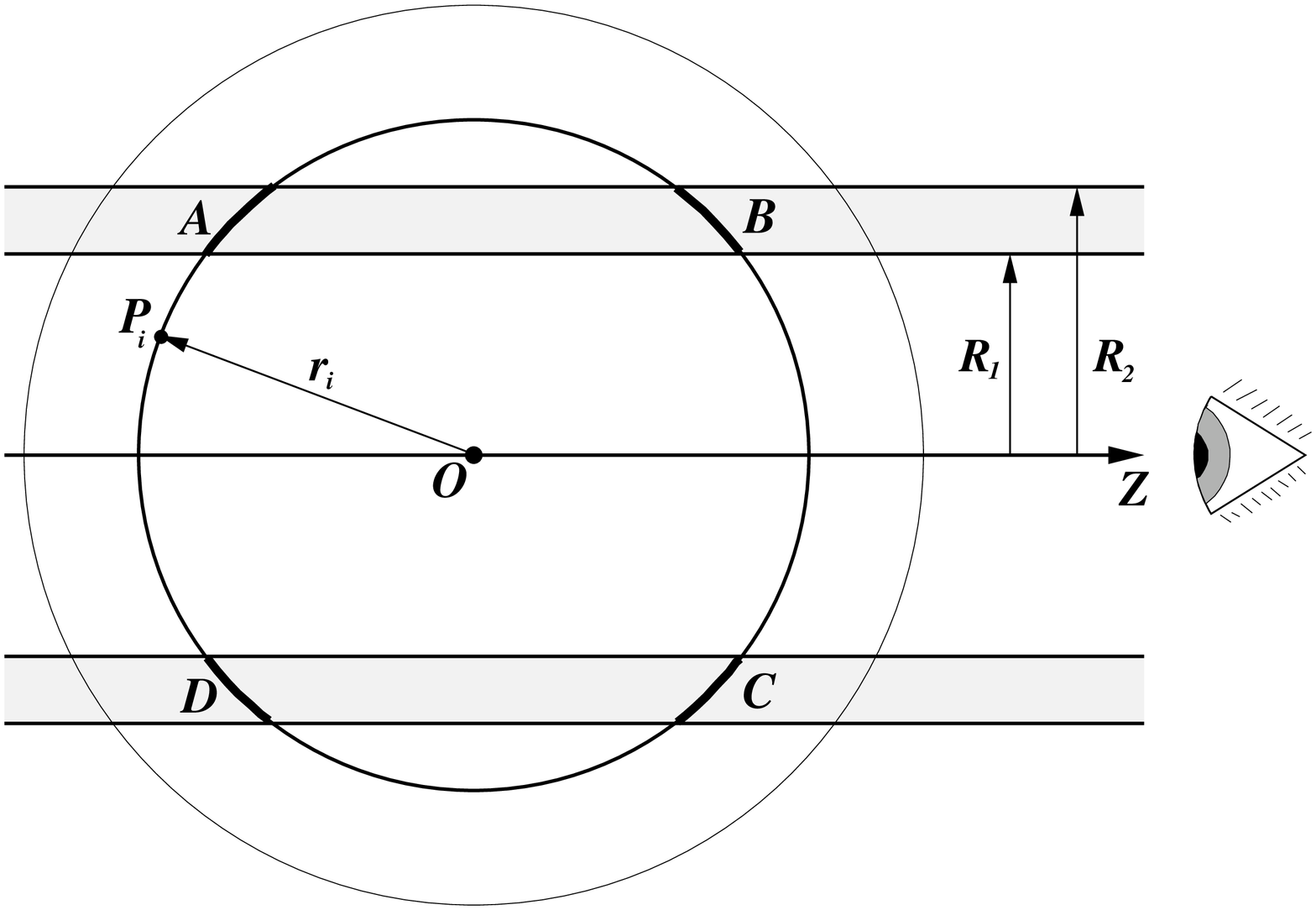}
\caption {Scheme explaining the projection procedure. The stellar cluster is marked
with its center $O$ and thin line circular boundary.  The axis $Z$ is
arbitrarily chosen to point toward observer. We show a sphere containing the
$i$-th particle $P_i$, located at the distance $r_i$ from the center of the
cluster, as a thick line circle. The range of projected distances $R_1\dots R_2$
is shown as two vertical shadowed bars. In 3D, it corresponds to space between
two concentric cylinders of radii $R_1$ and $R_2$ with the axis $OZ$ as their
central axis. These two cylinders cut segments out of the sphere containing the
particle $P_i$, which are marked as four regions $A$, $B$, $C$, and $D$. In 3D,
these four regions correspond to two segments of the sphere.  For an arbitrary
rotation of the cluster around its center $O$, the probability to find the
particle $P_i$ within the bin $R_1\dots R_2$ is equal to the ratio of the
surface area of the segments to the total surface area of the sphere $4\pi
r_i^2$.
\label{scheme} }
\end{figure}


Let us consider an $i$-th stellar particle $P_i$ in a spherically symmetric
cluster, located at the distance $r_i$ from the center of the cluster
(Figure~\ref{scheme}). We want to find its averaged contribution to one
projection bin, corresponding to the range of projected distances $R_1\dots
R_2$.  For an arbitrarily rotated cluster, the probability $w_i$ to detect our
particle inside the bin (the shaded areas in Figure~\ref{scheme}) is equal to
the fraction of the surface area of the sphere, containing the particle $P_i$,
cut out by two concentric cylinders with the radii $R_1$ and $R_2$. It is easy
to show that this probability is

\begin{equation}
w_i=\cases{
0, & $r_i<R_1$;\cr
\left[1-(R_1 / r_i)^2\right]^{1/2}, & $R_1\leqslant r_i<R_2$;\cr
\left[1-(R_1 / r_i)^2 \right]^{1/2}-\left[1-(R_2/r_i)^2 \right]^{1/2}, & $r_i\geqslant R_2$.\cr
}
\label{w_i}
\end{equation}

\noindent The projected surface mass density $\zeta$
for the bin $R_1\dots R_2$ averaged over all possible rotations of the cluster
is

\begin{equation}
\zeta(R_1,R_2)=\frac{1}{\pi (R_2^2-R_1^2)}\sum_{i=1}^N m_i w_i,
\label{zeta}
\end{equation}

\noindent where $N$ is the total number of stellar particles in the cluster and
$m_i$ is the mass of the $i$-th particle.  The averaged value of a projected scalar
quantity $Q$ is

\begin{equation}
Q(R_1,R_2)= \sum_{i=1}^N w_i Q_i \; \left(\sum_{i=1}^N w_i\right)^{-1},
\label{eqQ}
\end{equation}

\noindent so the probability $w_i$ serves as a weight for scalar variables. 
For example, to calculate the square of the projected three-dimensional velocity
dispersion $\sigma_{3D}$, corresponding to the projected distances range
$R_1\dots R_2$, one has to set $Q_i=(V_{x,i}^2+V_{y,i}^2+V_{z,i}^2)$ in
equation~(\ref{eqQ}), where $(V_{x,i},V_{y,i},V_{z,i})$ are the velocity
components of the $i$-th particle. Observationally, one can determine
$\sigma_{3D}$ by measuring both line-of-sight velocities and two proper motion
components for stars located within the interval $R_1\dots R_2$ of projected
radial distances.

One can show that the above method is superior to a straightforward projection
onto one plane or three orthogonal planes by considering a contribution made by
one particle to the bin $R_1\dots R_2$. In case of the straightforward
projection, there is a high probability that the particle will make no
contribution to the projection bin. In our method, every particle with $r_i>R_1$
will make a contribution to the bin, which will make the radial profile of the
projected quantity significantly less noisy. Central value of the projected
quantity is obtained by setting $R_1$ to zero in
equations~(\ref{w_i})--(\ref{eqQ}). In this case, all particles make
contribution to the averaged value of the quantity.

The situation with vector quantities is a bit more complicated. We designed a
simplified procedure which is based on the assumption that the bin sizes are
small (so the number of bins is large).  Each particle with an associated vector
quantity (say, velocity vector) is first rotated around the $OZ$ axis to
bring the particle in the plane $XOZ$. Next, the particle is rotated around the
axis $OY$ to make the projected radial distance of the particle equal to the
radial distance of the center of the bin $R\equiv (R_1+R_2)/2$. (The observer is
assumed to be at the end of the axis $OZ$.)

The detailed procedure is as follows. For a particle with coordinates $(x,y,z)$,
after the two rotations the associated vector $(V_x,V_y,V_z)$ is transformed
into vector $(V''_x,V''_y,V''_z)$ with the following components:

\begin{equation}
V''_x=V'_{xz} \sin \psi', \quad V''_y=V_{xy} \sin \varphi', \quad V''_z=V'_{xz} \cos \psi'.
\end{equation}

\noindent The parameters $V'_{xz}$, $V_{xy}$, $\psi'$, and $\varphi'$ can be obtained
after a series of the following calculations:

\begin{equation}
\begin{array}{llll}
r\equiv (x^2+y^2+z^2)^{1/2}; & r_{xy}\equiv (x^2+y^2)^{1/2}; & V_{xy}\equiv(V_x^2+V_y^2)^{1/2}; &\\
\sin\varphi=V_y/V_{xy}; & \cos\varphi=V_x/V_{xy}; & \sin\alpha=y/r_{xy}; & \cos\alpha=x/r_{xy};\\
\varphi'\equiv\varphi-\alpha; &&&\\
V'_{xz}\equiv (V_{xy}^2 \cos^2\varphi'+V_z^2)^{1/2}; &&&\\
\sin\psi=V_{xy} \cos\varphi'/V'_{xz}; & \cos\psi=V_{z}/V'_{xz};  & \sin\theta=r_{xy}/r; & \cos\theta=z/r;\\
\psi'\equiv\psi-\theta+\arcsin\varkappa.&&&\\
\end{array}
\end{equation}

\noindent Here $\varkappa$ is equal to $R/r$ if $R/r< 1$ and is equal to 1 otherwise;
$r$, $r_{xy}$, $V_{xy}$, $\varphi$, $\alpha$, $\theta$, and $\psi$ are
intermediate variables.

If vector $(V_x,V_y,V_z)$ is velocity vector, then substituting $V''^2_z$
in place of $Q_i$ in equation~(\ref{eqQ}) will give us the value of the square of
the line-of-sight velocity dispersion $\sigma_r$, averaged over all possible rotations of the
cluster, for the bin $R_1\dots R_2$. Similarly, replacing $Q_i$ with $V''^2_x$ and
$V''^2_y$ will produce the averaged values of the square of the radial and
tangential components of the proper motion dispersion, respectively.

\end{appendix}

\acknowledgements We would like to thank Volker Springel for his help with introducing
an external static potential in GADGET. S. M. is partially supported by
SHARCNet. The simulations reported in this paper were carried out on McKenzie
cluster at the Canadian Institute for Theoretical Astrophysics.


\end{document}